\documentclass[12pt]{article}
\usepackage{amssymb}
\usepackage{soul}
\usepackage{amsmath}
\usepackage{amsthm}
\usepackage{latexsym}
\usepackage{verbatim}
\usepackage{epsfig}

\usepackage{amsbsy}
\usepackage{amscd}
\usepackage{bm}
\usepackage{graphicx,psfrag,epsf}
\usepackage{enumerate}
\usepackage[numbers,sort]{natbib}
\usepackage{bm}
\usepackage{color}
\usepackage{nicefrac}
\newcommand\bsfrac[2]{%
\scalebox{-1}[1]{\nicefrac{\scalebox{-1}[1]{$#1$}}{\scalebox{-1}[1]{$#2$}}}%
}

\newcommand{\blind}{0}

\providecommand{\M}[1]{\mathbf#1}
\providecommand{\mc}[1]{\mathcal#1}
\providecommand{\mc}[1]{\mathcal#1}

\newcommand{\R}{{\mathbb R}}

\DeclareMathOperator{\E}{\mathbf{E}}
\DeclareMathOperator{\p}{\mathbf{P}}

\DeclareMathOperator{\var}{Var}
\DeclareMathOperator{\cov}{Cov}

\providecommand{\T}{\top} 

\DeclareMathOperator*{\argmin}{argmin}

\providecommand{\wt}[1]{\widetilde{#1}}
\providecommand{\wh}[1]{\widehat{#1}}

\providecommand{\nnorm}[1]{ \lVert#1 \rVert}

\newcommand{\nscp}[2]{\langle#1, #2\rangle}

\newcommand{\dev}[2]{\Big\lvert _{{#1}={#2}}}

\newcommand{\blanco}[1]{  }

\newcommand{\deriv}[3]{%
\ifthenelse{#1 = 1}{\frac{d\,#2}{d\,#3}}{\frac{d^{{#1}} #2}{d{#3}^{{#1}}}}
} 

\newcommand{\sss}[2]{\overset{\mbox{{\normalsize #1}}}{\mbox{{\scriptsize #2}}}}

\newcommand{\partials}[3]{%
\ifthenelse{#1 = 1}{\frac{\partial\,#2}{\partial\,#3}}{\frac{\partial^{#1}
    #2}{\partial#3^{#1}}}
} 

\def\su{\sum_{i=1}^n}

\def \coloneq{\mathrel{\mathop:}=}

\def \eps{\varepsilon}

\newtheorem{theo}{Theorem}
\newtheorem{propo}{Theorem}

\newtheorem{theorem}[theo]{Theorem}

  \newtheorem{prop}[propo]{Proposition}

\begin{document}

\def\spacingset#1{\renewcommand{\baselinestretch}%
{#1}\small\normalsize} \spacingset{1}


\if0\blind
{
  \title{\bf A Pseudo-Likelihood Approach to Linear Regression with Partially Shuffled Data}
  \author{Martin Slawski\thanks{
    Partially supported by NSF award CRII: CIF: 1849876} $\qquad$ Guoqing Diao\hspace{.2cm}\\
    Department of Statistics, George Mason University\\
    and \\
    Emanuel Ben-David \\
    U.S.~Census, CSRM}
  \maketitle
} \fi

\if1\blind
{
  \bigskip
  \bigskip
  \bigskip
  \begin{center}
    {\LARGE\bf Title}
\end{center}
  \medskip
} \fi

\bigskip
\begin{abstract} Recently, there has been significant interest in linear regression in the situation
where predictors and responses are not observed in matching pairs corresponding to the same statistical unit as a consequence of separate data collection and uncertainty in data integration. Mismatched pairs can considerably impact the model fit and disrupt the estimation of regression parameters. In this paper, we present a method to adjust for such mismatches under ``partial shuffling" in which a sufficiently large fraction of (predictors, response)-pairs are observed in their correct correspondence. The proposed approach is based on a pseudo-likelihood in which each term takes
the form of a two-component mixture density. Expectation-Maximization schemes are proposed 
for optimization, which (i) scale favorably in the number of samples, and (ii) achieve excellent statistical performance relative to an oracle that has access to the correct pairings as certified by simulations and case studies. In particular, the proposed approach can tolerate considerably larger fraction of mismatches than existing approaches, and enables estimation of the noise level as well as the fraction of mismatches. Inference for the resulting estimator (standard errors, confidence intervals) can be based on established theory for composite likelihood estimation. Along the way, we also propose a statistical test for the presence of mismatches and establish its consistency under suitable conditions.  
\end{abstract}

\noindent%

\spacingset{1.45}
\section{Introduction}\label{sec:intro}
A tacit assumption in linear regression is that each of the (predictors, response)-pairs $\{(\M{x}_i, y_i) \}_{i = 1}^n$ is associated with the same underlying statistical unit. However, there are scenarios in which 
the $\{ \M{x}_i \}_{i = 1}^n$ and the $\{ y_i \}_{i = 1}^n$ were collected
separately, and there is uncertainty about which of the pairs $\{ (\M{x}_i, y_j)\}_{i < j}$ are in correspondence to each other. Pioneering work by
DeGroot, Goel, and collaborators \cite{DeGroot1971, DeGroot1976, Goel1975, Goel1987, DeGroot1980} has formalized this setting under the notion of ``Broken Sample": it is assumed that $\{ (\M{x}_{\pi^*(i)}, y_i) \}_{i = 1}^n$ are i.i.d.~observations from
some joint distribution $P_{\M{x},y;\theta^*}$, where $\theta^*$ is an unknown parameter and $\pi^*$ is an unknown permutation of $\{1,\ldots,n\}$. To give an example, $P_{\M{x},y;\theta^*}$ might be a Gaussian distribution with $\theta^*$ representing the mean and covariance matrix. Depending on the problem, inference for both $\theta^*$ and $\pi^*$ can be of interest.

More recently, there has been a surge of interest in the above setup in the context of linear regression, driven by applications in engineering and promising new developments in the mathematical signal processing and machine learning literature. Specifically, the following model has been studied under the terms ``Unlabeled Sensing" \cite{Unnikrishnan2015}, ``Regression with unknown permutation" \cite{Pananjady2016,Emiya2014}, and ``Regression with shuffled data" \cite{Abid2017, Hsu2017}: 
\begin{align}\label{eq:limo_perm}
\begin{split}
&y_i = \M{x}_{\pi^*(i)}^{\T} \beta^* + \sigma_* \eps_i, \quad i=1,\ldots,n,\\ \Longleftrightarrow \; \; &\M{y} = \Pi^* \M{X} \beta^* + \sigma \bm{\epsilon}, \; \; 
\M{y} = (y_i)_{i = 1}^n, \; \Pi^* =  \left(I(\pi^*(i) = j) \right)_{i,j=1}^n, \; \M{X}^{\T} = [\M{x}_1 \ldots \M{x}_n], \; \bm{\epsilon} = (\eps_i)_{i = 1}^n. 
\end{split}
\end{align}
In \eqref{eq:limo_perm}, $\beta^* \in \R^d$ denotes the regression parameter, the $\{ \eps_i \}_{i = 1}^n$ represent i.i.d.~zero-mean and unit-variance errors, and $\sigma_*$ is referred to as ``noise level". Model \eqref{eq:limo_perm} has been considered from the point of view of signal recovery (with $\beta^*$ representing an unknown signal of interest) based on (noisy) linear sensing at the $\{ \M{x}_i \}_{i = 1}^n$, with the additional caveat that those linear measurements are received in an unknown
order. For example, each measurement may come with an inaccurate time stamp, and 
as result, measurements are received in a shuffled order \cite{Balakrishnan1962}. Specific applications of \eqref{eq:limo_perm} in signal processing and sensors networks are reviewed in \cite{Unnikrishnan2013, Unnikrishnan2015, Pananjady2016, Pananjady2017, Haghighatshoar2017}. 

Another important domain in which model \eqref{eq:limo_perm} is of interest is data integration. Specifically, consider two data files $A$ and $B$, with $A$ containing
the response variable $y$ and $B$ containing predictor variables $\M{x}$ for a
set of statistical units common to $A$ and $B$. For example, $A$ may contain the
annual income of a set of individuals, while $B$ may contain a collection of 
demographic variables, and the goal is to fit a linear regression model for income
based on those variables. To perform this task, file $A$ needs to be merged with file $B$, which is straightforward only if both files are equipped with unique identifiers. However, it is common that the data analyst does not have access to
such identifiers, e.g., because of privacy concerns. In this case, linkage of $A$ and $B$ needs to be based on a combination of variables that are contained in both files (so-called matching variables), with the possibility of
ambiguities and the potential for \emph{linkage error}, i.e., a record in $A$ is 
not matched to the correct counterpart in $B$. Therefore, model \eqref{eq:limo_perm}
can be used to account for errors (mismatches) in post-linkage regression analysis, a long-standing problem in statistics dating back to the work in \cite{Neter65} that is particularly relevant in the area of official statistics and the work of government agencies like the U.S.~Census Bureau \cite{Scheuren93, Scheuren97}. The latter regularly combines data from a variety of sources such as administrative data, sample survey, and census data. The main purpose of combining data is to reuse existing data, reduce the cost of data collection, research, and the burden on responders. In spite of its relevance to this application, model \eqref{eq:limo_perm} has hardly been considered directly. Instead, the common
approach is to use information about the linkage process, e.g., the probability
of a mismatch given a certain configuration for the matching variables, to
construct suitable estimators that curb the impact of linkage error on the
regression fit \cite{Lahiri05, Chambers2009, Hof12, Hof2014}. However, information about the linkage process may be scarce or unavailable, in which case it can be useful to resort to \eqref{eq:limo_perm}. 

\noindent \emph{Related Work}. Several recent papers have studied estimation of $\beta^*$ and/or $\pi^*$ under model \eqref{eq:limo_perm}, predominantly under
random Gaussian design with $\{ \M{x}_i \}_{i = 1}^n \overset{\text{i.i.d.}}{\sim} \, N(0, I_d)$ and Gaussian errors. The paper \cite{Unnikrishnan2015} shows that in the noiseless case ($\sigma_* = 0$), $\beta^*$ can be  
uniquely recovered by exhaustive search over permutations if $n > 2d$. Regarding the noisy case, a series of properties have been established
for the least squares problem 
\begin{equation}\label{eq:ls_permutation}
 \min_{ \Pi \in \mc{P}_n, \, \beta \in \R^d} \, \nnorm{\M{y} -  \Pi \mathbf{X} \beta}_2^2,
\end{equation}
where $\mc{P}_n$ denotes the set of $n$-by-$n$ permutation matrices. Problem \eqref{eq:ls_permutation} is a specific quadratic assignment problem \cite{Burkard2009}. A result in
\cite{Pananjady2016} shows that \eqref{eq:ls_permutation} is NP-hard. The paper \cite{Pananjady2016} also derives necessary and sufficient conditions for exact and approximate recovery of $\Pi^*$ based on \eqref{eq:ls_permutation}, and elaborates on the significance of the signal-to-noise ratio (SNR) $\nnorm{\beta^*}_2^2 / \sigma^2$ in this context. An excessively large SNR of the order $n^2$ is proved to be a necessary condition for approximate permutation recovery for \emph{any} estimator. In a similar spirit, the work \cite{Hsu2017} shows that the SNR needs to be at least of the order $d / \log \log n$ to ensure approximate recovery of $\beta^*$. In fact, problem \eqref{eq:ls_permutation} can be shown to suffer from overfitting due to the extra freedom in optimizing $\Pi$ \cite{Abid2017, SlawskiBenDavid2017}.

Tractable algorithms for \eqref{eq:ls_permutation} with provable guarantees are scarce at this point: the scheme in \cite{Hsu2017} has polynomial time complexity, but is ``not meant for practical deployment'' as the authors state themselves. The convex relaxation of \eqref{eq:ls_permutation} in which $\mc{P}_n$ is replaced by its convex hull, the set of doubly stochastic matrices, was observed to yield poor performance \cite{Emiya2014}. The works \cite{Wu1998, Abid2018} discuss alternating minimization as well as the use of the Expectation-Maximization (EM) Algorithm (combined with MCMC sampling) in which $\Pi^*$ constitutes missing data. The recent paper \cite{Tsakiris2018b} discusses a branch-and-bound scheme with promising empirical performance on small data sets; the theoretical properties of the approaches \cite{Wu1998, Abid2018, Tsakiris2018b} remain to be investigated. 

In view of the aforementioned computational and statistical barriers, the paper \cite{SlawskiBenDavid2017} discusses a simplified setting of \eqref{eq:limo_perm} in which $\pi^*(i) = i$ except for at most $k \ll n$ elements of $\{1,\ldots,n\}$; $\pi^*$ is called $k$-\emph{sparse} in this case. The authors of \cite{SlawskiBenDavid2017} show that under this restriction on $\pi^*$, the constrained least squares estimator corresponding to \eqref{eq:ls_permutation} has
desirable statistical properties if the fraction $k/n$ is not too large. Moreover, a convex relaxation of that constrained least squares problem yields an estimator of $\beta^*$ that is consistent under suitable conditions on $k/n$; the permutation can be estimated subsequently by sorting (cf.~Eq.~\eqref{eq:plugin} below). The papers \cite{SlawskiRahmaniLi2018, SlawskiBenDavidLi2019} consider extensions of the approach taken in \cite{SlawskiBenDavid2017} to a multivariate regression setup in which the $\{ y_i \}_{i = 1}^n$ have dimension $m \geq 1$ each. It is shown that permutation recovery, i.e., the event $\{ \wh{\Pi} = \Pi^* \}$ for a suitable estimator $\wh{\Pi}$ of $\Pi^*$, can succeed without stringent conditions on the signal-to-noise ratio once $m$ is at least of the order $\log n$. The paper \cite{ZhangSlawskiLi2019} complements this result with matching information-theoretic lower bounds. Motivated by applications in automatic term translation, the paper \cite{Shi2018} considers a closely related setup which the authors term ``spherical regression with mismatch corruption" with responses and predictors being contained in the unit sphere of in $\R^m$ (in \cite{Shi2018}, $m = d$). In addition to sparsity, the paper \cite{Shi2018} additionally assumes $\Pi^*$ to have a block structure. On the other hand, in \cite{Shi2018} $\Pi^*$ is not required to be a permutation to allow a slightly more general class of mismatches. 

Finally, it is worth mentioning that instead of linear regression, the papers \cite{Carpentier2016, Weed2018} study isotonic regression, i.e., $y_i = f^*(x_{\pi^*(i)}) + \sigma_* \eps_i$, $1 \leq i \leq n$, with $\{ x_i \}_{i = 1}^n$ and $\{ y_i \}_{i = 1}^n$ being one-dimensional, and a non-decreasing regression function $f^*$. 


\noindent \emph{Contributions}. In this paper, we build on the setup of sparse permutations as put forth in \cite{SlawskiBenDavid2017}. The approach proposed herein improves 
over the approach in \cite{SlawskiBenDavid2017} with regard to two important
aspects. One of the downsides in \cite{SlawskiBenDavid2017} is that mismatches
are treated as generic data contamination, which leads to a substantial loss of information. As a result, performance degrades severely as the fraction 
of mismatches $k/n$ increases. A second drawback of the approach is its dependence
on a tuning parameter involving the noise level $\sigma_*$, which is generally not known nor easy to estimate. By contrast, the approach proposed herein is in principle tuning-free (apart from the choice of a suitable initial solution), and produces estimates of 
the noise level $\sigma_*$ and the fraction of mismatches $\alpha_* = k/n$ in addition 
to an estimate of the regression parameter. Moreover, the approach is far less affected
as $\alpha_*$ increases and empirically performs rather closely to the oracle least squares estimator equipped with knowledge of $\pi^*$ (cf.~$\S$\ref{sec:verify}); while in theory, an arbitrary constant fraction of mismatches can be tolerated, $\alpha_* \approx 0.7$ typically constitutes the limit in practice. In addition, the proposed approach also avoids a quadratic runtime in $n$ that is incurred for alternatives such as the Lahiri-Larsen estimator \cite{Lahiri05}. Optimization is based on a pseudo-likelihood 
having the form of a likelihood for fitting a two-component mixture model, with one component 
corresponding to a regular linear regression model without mismatches and the second component
accounting for extra variation due to mismatches. Despite the nonconvexity of the resulting
optimization problem, reasonable approximate solutions can be obtained via the EM algorithm and 
one of its variants \cite{Lange1995, Titterington1984} whose initialization is discussed in detail in $\S$\ref{subsec:init}. The EM schemes are easy to implement and exhibit only a linear dependence in $n$. By leveraging well-developed theory on composite likelihood estimation, asymptotic standard errors 
and confidence intervals for $(\beta^*, \sigma_*, \alpha_*)$ can be obtained (cf.~Theorem \ref{theo:asymptotics}). Along the way, we also propose a test for the null hypothesis 
$H_0: \Pi^* = I_n$, i.e., a test for the presence of mismatches, and show its consistency
under suitable conditions ($\S$\ref{subsec:hypotest}). 

\section{Approach}\label{sec:meth}
To formally introduce the approach proposed herein, we make the following assumptions. 
\begin{itemize}
\item[(A1)] The permutation $\pi^*$ is assumed to be chosen uniformly
      at random from the set of permutations $\{ \pi: \; \su I(\pi(i) \neq  i) = k  \}$
      moving exactly $k$ indices of $\{1,\ldots,n\}$.  
\item[(A2)] Conditional on $\pi^*$, the pairs  $\{ (\M{x}_{\pi^*(i)}, y_i) \}_{i = 1}^n$ are i.i.d.~zero-mean random variables, drawn        from a joint distribution with density $f_{\M{x}, y}(\M{x}, y) = f_{y|\M{x}}(y) \cdot f_{\M{x}}(\M{x})$ with $f_{y|\M{x}} \sim N(\M{x}^{\T} \beta^*, \sigma_*^2)$. 
\end{itemize}
Define indicator variables $z_i = I(\pi^*(i) \neq i)$, $i = 1,\ldots,n$, and fix
an arbitrary index $i \in \{1,\ldots,n \}$.
Under (A2), it then holds that 
\begin{equation}\label{eq:mixture_pieces}
y_i | \{\M{x}_i, z_i = 0\} \sim N(\M{x}_i^{\T} \beta^*, \sigma_*^2), \qquad \qquad 
y_i | \{\M{x}_i, z_i = 1\} \sim f_y, 
\end{equation}
where $f_y(y) = \int f_{y|\M{x}}(y) f_{\M{x}}(\M{x}) \; d\M{x}$. In fact, note that conditional on $\{ z_i = 1 \}$, $y_i$ is independent of
$\M{x}_i$, and as a result the conditional distribution coincides with the marginal distribution 
of $y$. In conclusion, (A1) and \eqref{eq:mixture_pieces} imply that $y_i|\M{x_i}$ follows a two-component mixture with proportions $1-\alpha_*$ and $\alpha_* = k/n$, i.e., 
\begin{equation}\label{eq:mixture_general}
y_i | \M{x}_i \sim (1 - \alpha_*) N(\M{x}_i^{\T} \beta^*, \sigma_*^2) +  \alpha_* f_y   
\end{equation}
\emph{Remarks.} (i) Assumption (A1) can be considerably relaxed without affecting \eqref{eq:mixture_general}. Specifically, it suffices to assume that the indicators $\{ z_i \}_{i = 1}^n$ are independent 
of $\{(\M{x}_i, y_i) \}_{i = 1}^n$ and satisfy $\p(z_i) = \alpha_*$, $1 \leq i \leq n$. In fact, the latter does not even require $\pi^*$ to be a permutation of $\{1,\ldots,n\}$.\\ (ii) The zero-mean assumption in (A2) is merely made for convenience as it eliminates the need for an intercept. 
\vskip1ex
Since estimation of the marginal density in $f_y$ \eqref{eq:mixture_general} can be performed based on the $\{ y_i \}_{i = 1}^n$ only and is thus not affected by the presence of 
$\pi^*$, $f_y$ can be assumed to be effectively known given that $n$ is sufficiently large and can consequently be estimated with small error, be it in a parametric (e.g., by assuming 
$f_y \sim N(0, \tau_*^2)$) or in nonparametric fashion (by density estimation). 

In the sequel, we focus on isotropic Gaussian design as considered in \cite{Pananjady2016, SlawskiBenDavid2017}, i.e., $f_{\M{x}} \sim N(0, I_d)$. In this case, it holds that $f_y \sim N(0, \nnorm{\beta^*}_2^2 + \sigma_*^2 )$ and accordingly 
\begin{equation}\label{eq:gaussian_mixture}
y_i | \M{x}_i \sim (1 - \alpha_*) N(\M{x}_i^{\T} \beta^*, \sigma_*^2) +  \alpha_* N(0, \nnorm{\beta^*}_2^2 + \sigma_*^2).  
\end{equation}
The above considerations suggest the following ``likelihood" approach 
\begin{equation}\label{eq:likelihood}
\max_{\beta \in \R^d, \, \sigma^2 > 0, \, \alpha \in [0,1] } \, \prod_{i = 1}^n f(y_i|\M{x}_i),  
\end{equation}
where $f(y_i | \M{x}_i)$ refers to the density of the above Gaussian mixture distribution, i.e.,   
{\small \begin{equation}\label{eq:GMM}
 f(y_i|\M{x}_i) =   \frac{1- \alpha}{\sqrt{2 \pi \sigma^2}}\exp \left(-\frac{(y_i - \M{x}_i^{\T} \beta)^2}{2 \sigma^2} \right)  
+  \frac{\alpha}{\sqrt{2 \pi (\sigma^2 + \nnorm{\beta}_2^2)}} \exp \left(-\frac{y_i^2}{2  (\sigma^2 + \nnorm{\beta}_2^2)} \right), \; i=1,\ldots,n. 
\end{equation}}
The quotation marks above indicate that the objective in \eqref{eq:likelihood} is not a genuine likelihood function since 
$\{y_i | \M{x}_i \}_{i = 1}^n$ are not \emph{independent} observations from the Gaussian mixtures given in \eqref{eq:GMM}. However, we can still treat \eqref{eq:likelihood} within the framework of \emph{pseudo likelihood}, or more specifically \emph{composite likelihood}. The pseudo-likelihood \eqref{eq:likelihood} is composed of likelihoods of individual observations, which constitutes the most basic variant of a composite likelihood. Nevertheless, the approach enjoys several attractive
properties including asymptotic normality at the standard rate and a closed form expression for the asymptotic covariance matrix, while avoiding the computational barrier that is associated with the unknown permutation as elaborated in the introduction.
\vskip2ex
The asymptotic normality result is stated in Theorem \ref{theo:asymptotics} below. Denote by $\theta^*  = (\beta^*, \sigma_*^2, \alpha_*)$ the unknown parameter, which is supposed to be an interior point of $\R^d \times [0,\infty) \times [0,1]$. Let further $\ell_{\text{p}}(\theta) = \su \ell_{i,\text{p}}, \;\; \ell_{i,\text{p}} \coloneq -\log(f(y_i | \M{x}_i;\theta))$ denote 
the negative pseudo log-likelihood with $f(y_i|\M{x}_i;\theta)$ as in \eqref{eq:GMM}, $\theta = (\beta, \sigma^2, \alpha)$. The global minimizer
$\argmin_{\theta} \wh{\ell}_{\text{p}}(\theta)$ is denoted by $\wh{\theta}_n$.
Equipped with those definitions and the following assumption (A3), we can state the following result.
\begin{itemize}
\item[(A3)] The $\{ \ell_{i, \text{p}} \}_{i = 1}^n$ satisfy the regularity conditions specified in Theorem 5.23 or in Theorem 5.41 in \cite{vanderVaart1998}.  
\end{itemize}
\begin{theorem}\label{theo:asymptotics}
Under (A1), (A2), and (A3), $n^{1/2}(\widehat{\theta}_n - \theta^*) \rightarrow N(0, H_*^{-1} G^* H_*^{-1})$ in distribution as $n\rightarrow \infty$,  where $H_* = \M{E}\left[-\nabla_{\theta}^2 \log f(y|\M{x};\theta^*) \right]$ and  $G^* = \E \left[\nabla_{\theta} \log f(y|\M{x};\theta^*) \, \nabla_{\theta} \log f(y|\M{x};\theta^*)^{\T} \right]$. Here, $\nabla_{\theta}$ and $\nabla_{\theta}^2$ denote the gradient and Hessian with respect to $\theta$, respectively, $f(y|\M{x};\theta)$ denotes the density of a generic pair $(\M{x}, y)$ according to \eqref{eq:mixture_general}, and $\M{E}$ denotes the expectation
with respect to that density. Moreover, $H_*$ and $G^*$ can be  
consistently estimated by 
\begin{align}\label{eq:sandwich}
\wh{H} = n^{-1} \nabla_{\theta}^2 \ell_{\text{\emph{p}}}(\wh{\theta}_n),  \qquad \wh{G} = n^{-1} \su  \nabla_{\theta} \ell_{i,\text{\emph{p}}}(\wh{\theta}_n)  \, \nabla_{\theta} \ell_{i,\text{\emph{p}}}(\wh{\theta}_n)^{\T}.     
\end{align}
%
\end{theorem}
Theorem \ref{theo:asymptotics} can be proved by invoking well-established theory on composite likelihood theory; e.g., see \cite{Lindsay1988, Varin2011}. Hence we omit the details of the proof. Explicit expressions for the estimators $\wh{H}$ and $\wh{G}$, which are relevant in practice to estimate standard errors and to construct asymptotic confidence intervals, are provided in Appendix \ref{app:sandwich}.

\noindent \emph{Remark.} In this paper, we do not develop any novel approach for estimating the 
permutation $\pi^*$. If the latter is of interest, the plug-in approach in \cite{SlawskiBenDavid2017} can be applied. The latter is based on the optimization problem 
\begin{equation}\label{eq:plugin}
\min_{\Pi \in \mc{P}_{n}} \nnorm{\M{y} - \Pi \M{X} \wh{\beta}}_2^2 = -2\max_{\Pi \in \mc{P}_n} \nscp{\M{y}}{\M{X} \wh{\beta}} + c = -2\su y_{(i)} (\M{X} \wh{\beta})_{(i)} + c, 
\end{equation}
where $c = \nnorm{\M{y}}_2^2 + \nnorm{\M{X} \wh{\beta}}_2^2$ does not depend on $\Pi$ and the
subscript $(i)$ denotes the $i$-th order statistic, e.g., $y_{(1)} < \ldots < y_{(n)}$ (assuming the absence of ties). The relations in \eqref{eq:plugin} imply that for fixed $\beta$, the optimal permutation can be found by sorting $\{ y_i \}_{i = 1}^n$ and $\{ \M{x}_i^{\T} \beta \}_{i = 1}^n$. Statistical aspects of the plug-in approach \eqref{eq:plugin} are analyzed in \cite{SlawskiBenDavid2017} independent of specific properties of $\wh{\beta}$.

\subsection{Connection to Robust Regression}\label{subsec:robust}
\begin{figure}
\begin{centering}
\begin{tabular}{ccc}
\hspace*{-1.5ex}\includegraphics[width = 0.32\textwidth]{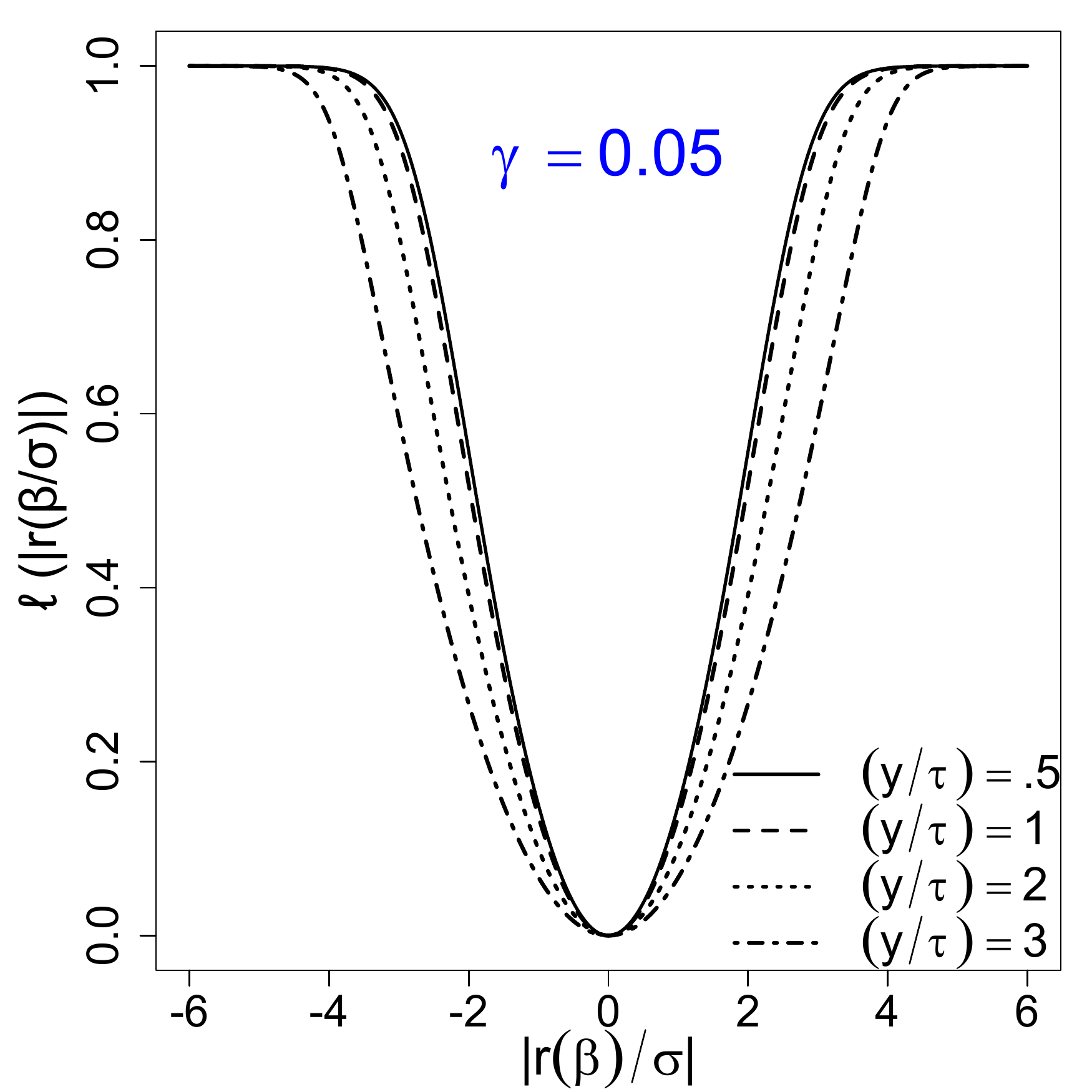} &
\hspace*{-.8ex}\includegraphics[width = 0.32\textwidth]{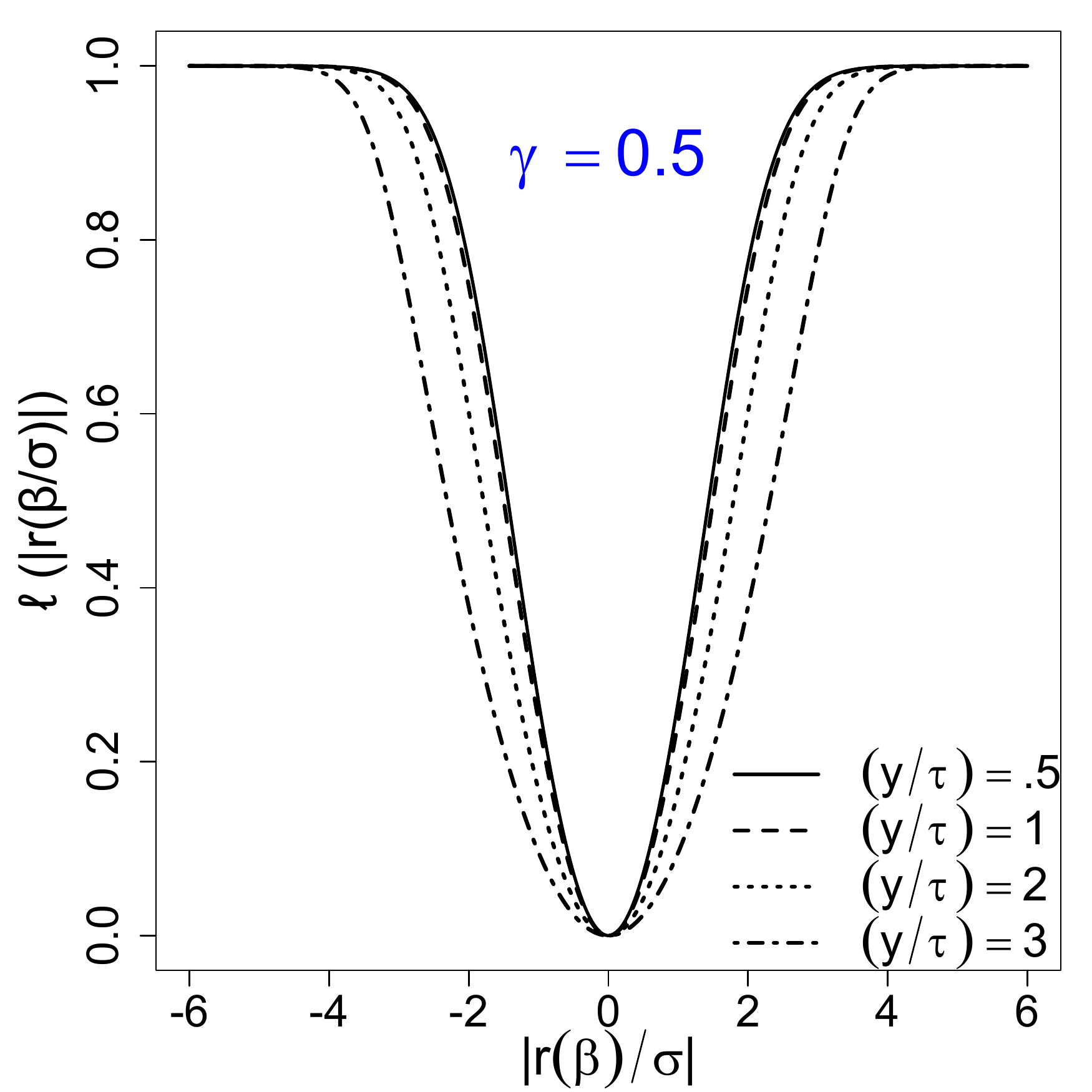} &
\hspace*{-.8ex}\includegraphics[width = 0.32\textwidth]{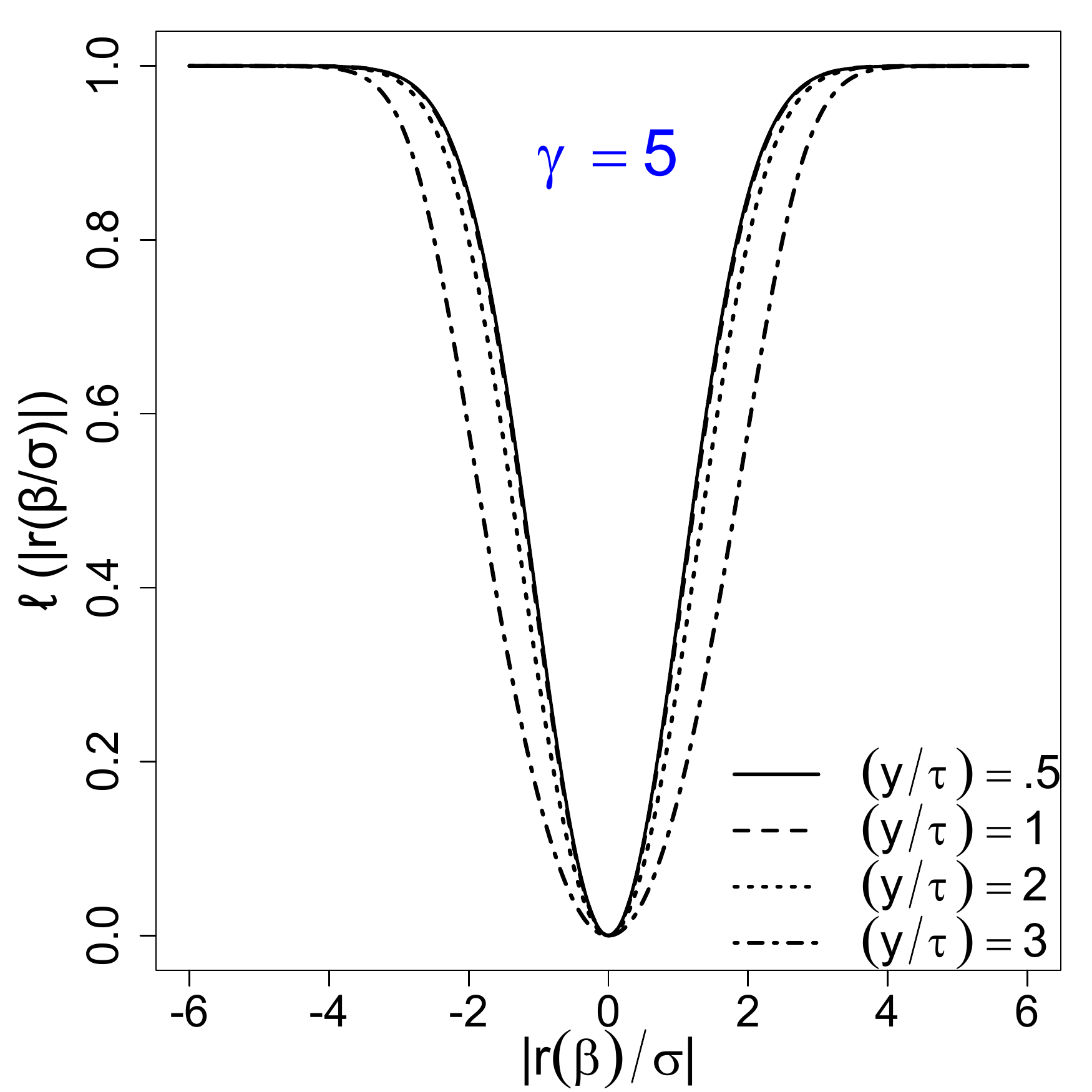}
\end{tabular}
\end{centering}    
\caption{Visualization of the loss function \eqref{eq:robustloss} for selected values 
of $\gamma$ and $y_i / \tau$.  }
\label{fig:robustestimation}
\end{figure}

The above pseudo-likelihood approach can be related to robust $M$-estimation as we elaborate below. This connection puts the present work in perspective with recent
prior work on the subject \cite{SlawskiBenDavid2017}. Consider the negative pseudo log-likelihood that follows from \eqref{eq:likelihood} and 
\eqref{eq:GMM}, up to additive constants: 
\begin{equation}\label{eq:logpseudolikelihood}
  \su -\log \left\{  \frac{1-\alpha}{\sigma} \exp \left(-\frac{(y_i - x_i^{\T} \beta)^2}{2 \sigma^2} \right)  + 
     \frac{\alpha}{\sqrt{\sigma^2 + \nnorm{\beta}_2^2}}
  \exp \left(-\frac{y_i^2}{2 (\sigma^2 + \nnorm{\beta}_2^2)} \right)  \right\}.
\end{equation}
With $\alpha$, $\sigma^2$ and $\tau = \sqrt{\sigma^2 + \nnorm{\beta}_2^2}$ considered as 
fixed, the above expression can be written as the following loss function $L(\beta)$, up to additive constants: 
\begin{align}\label{eq:robustloss}
L(\beta) = \su \ell\left(\left|\frac{r_i(\beta)}{\sigma} \right|;\, \gamma,\left| \frac{y_i}{\tau} \right| \right), \qquad \ell(z; \, a, b) &\coloneq -\log\left(\exp\left(-\frac{z^2}{2} \right)  + a \exp \left(-\frac{b^2}{2} \right) \right)\\
\gamma &\coloneq \frac{\alpha}{1-\alpha} \cdot \frac{\sigma}{\tau}. 
\end{align}
Figure \ref{fig:robustestimation} visualizes $\ell(\cdot; a, b)$ for selected values of $a$ and $b$; the function scales have been re-scaled to the range $[0,1]$. The shape of $\ell$ resembles a ``capped loss" such as Tukey's bisquare (e.g., \cite{Yohai2006}) commonly employed in robust regression. In fact, $\ell$ is uniformly bounded by  $-\log(a) + b^2/2$. 
For $\alpha = 0$, $\ell$ reduces to ordinary squared loss. As $\alpha$ increases,
$\ell$ levels off more quickly, and behaves like an indicator loss in the limit
$\alpha \rightarrow 1$. 
The above connection also highlights the advantages of the pseudo-likelihood approach compared to a plain robust $M$-estimation approach. The pseudo-likelihood can be
interpreted as the combination of observation-specific and self-calibrated robust
losses, where ``calibration" refers to tuning parameters that control the robustness vs.~efficiency trade-off (more informally speaking, parameters that control the range in which the loss function levels off). Moreover, the pseudo-likelihood integrates estimation of the parameters $\alpha_*$ and $\sigma_*^{2}$ of potential interest.


\subsection{Testing for the presence of mismatches}\label{subsec:hypotest}
\begin{figure}
\mbox{\hspace*{8ex}\includegraphics[width = 0.35\textwidth]{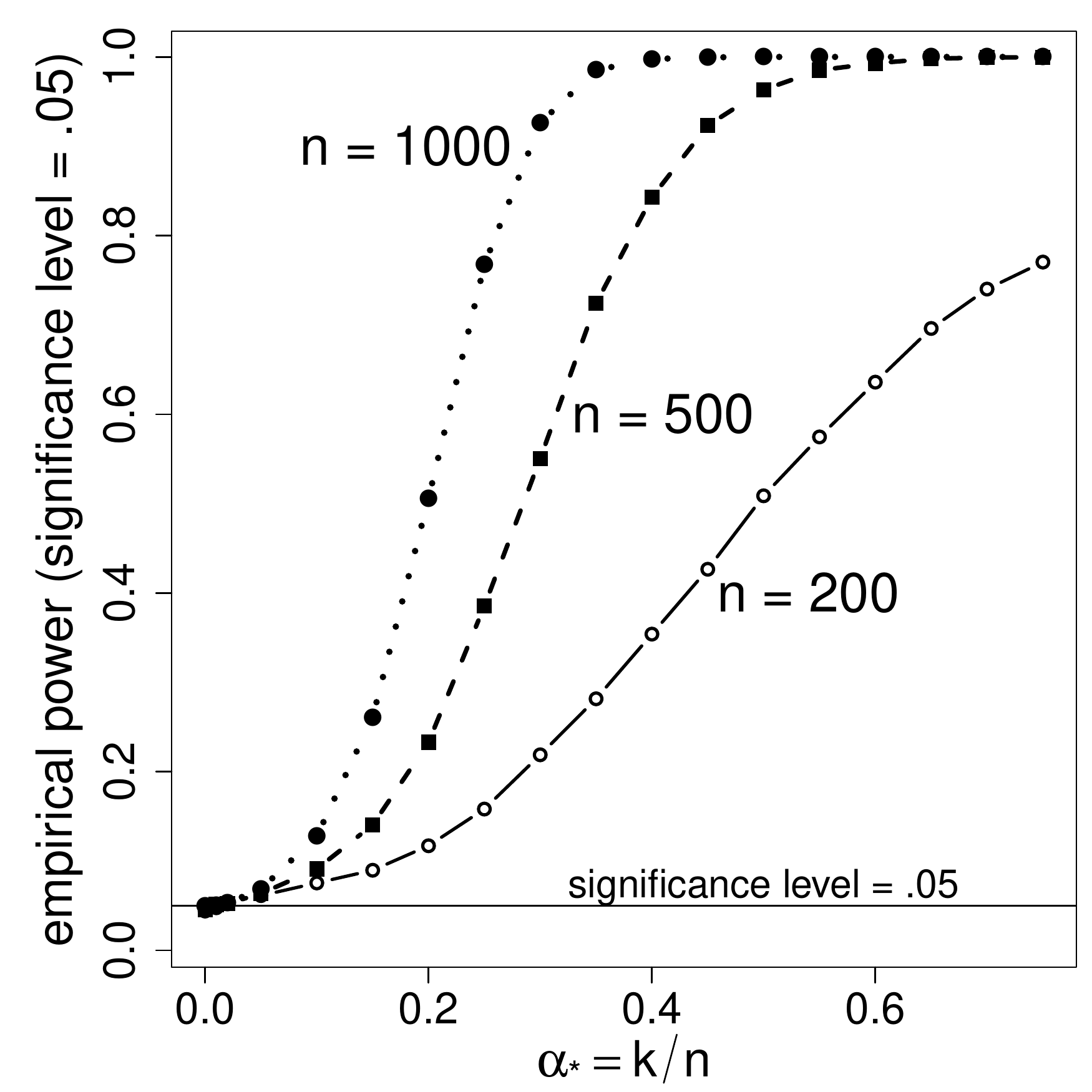}
\hspace*{10ex}\includegraphics[width = 0.35\textwidth]{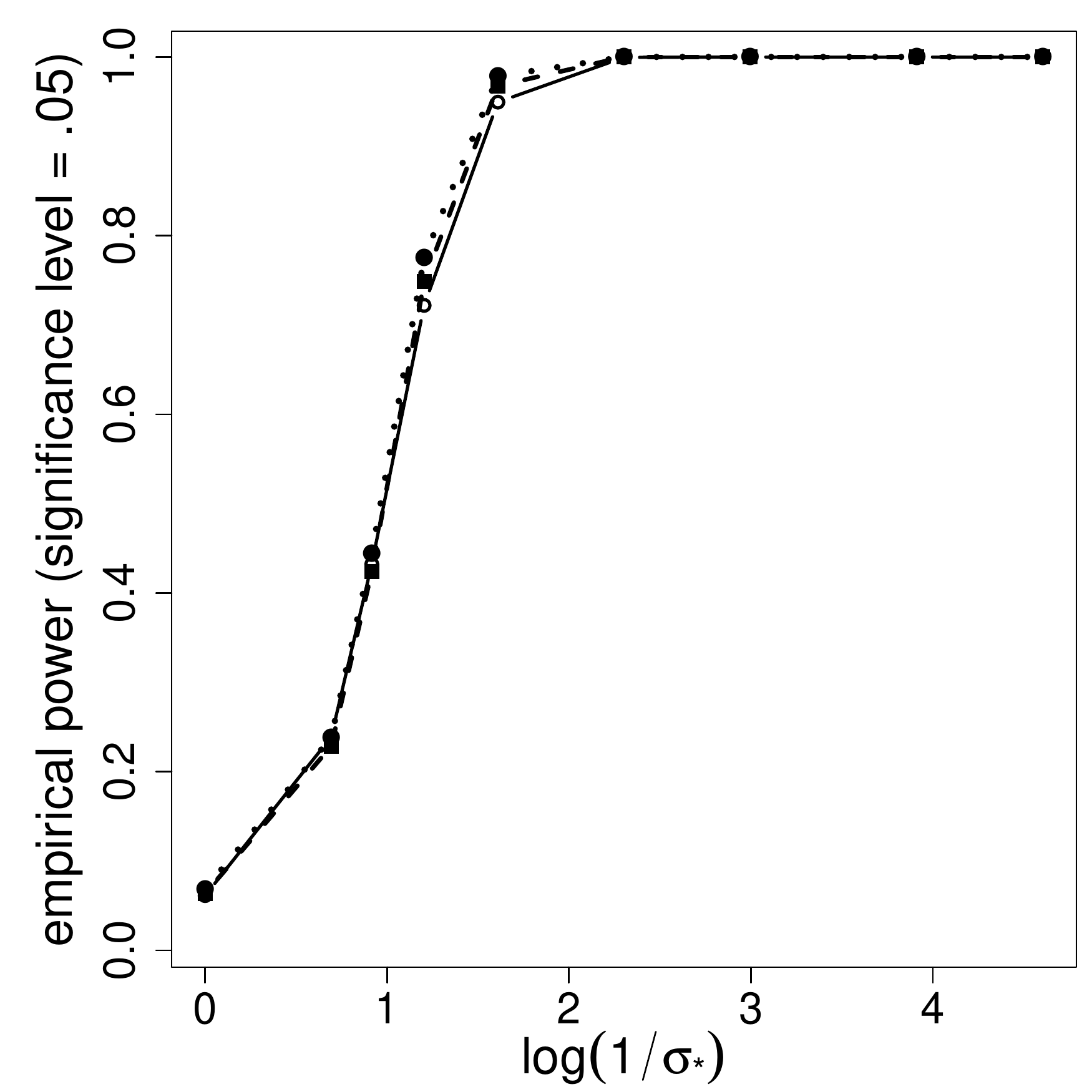}}
\caption{Empirical power of the CM test statistic (significance level .05) based on 10,000 replications from model \eqref{eq:limo_perm_gauss} under assumption (A1) with $\beta^*$ ($d = 10$) drawn uniformly at random from the unit sphere in dependence of $n \in \{200, 500, 1000\}$, $\alpha_* = k/n$ and $\sigma_*$. Left: empirical power for varying $\alpha_*$ and fixed $\sigma_* = 1$. Right: empirical power for varying $\sigma_*$ and fixed $\alpha_* = 0.05$.}\label{fig:power_gof}
\end{figure}

Before applying the approach developed at the beginning of this section, it is appropriate to test 
for the presence of mismatches. We here consider a statistical test for the hypothesis $H_0: \; \Pi^* = I_n$, or equivalently $H_0: \alpha_* = 0$. While 
one possible direction is the formulation of this test within the setting of mixture models \cite{Chen2009, Zhu2004}, a much more straightforward test can be performed based on the residuals $\wh{\bm{\epsilon}} = \texttt{P}_{\M{X}}^{\perp} \M{y}$ of the ordinary
least squares fit, where  $\texttt{P}_{\M{X}}^{\perp}$ denotes the orthoprojector on the orthogonal complement $\mc{U}$ of the column space
of $\M{X}$. Letting $\M{U}$ denote the $n$-by-$(n-d)$ matrix whose columns form an orthonormal basis of $\mc{U}$, we have 
$\wh{\bm{\epsilon}} = \texttt{P}_{\M{X}}^{\perp} \M{y} = \M{U} \M{U}^{\T} \M{y}$ and thus $\xi = \M{U}^{\T} \wh{\bm{\epsilon}} = \M{U}^{\T} \M{y}$. This yields
\begin{align}
    \xi =  \M{U}^{\T} \M{y} = \M{U}^{\T} \Pi^* \M{X} \beta^* + \sigma_* \M{U}^{\T} \bm{\epsilon} &=  \M{U}^{\T} \Pi^* \M{X} \beta^* + \zeta,\label{eq:derivation_test}\\ 
    &\overset{H_0}{=} \zeta \quad \text{where} \; \, \zeta \sim N(0, \sigma_*^2 I_{n-d}) \notag. 
\end{align}
This suggests the following statistical test (assuming that $\sigma_*$ is known): 
\begin{enumerate}
\item Compute $\xi = \M{U}^{\T} \M{y}$ via a singular value decomposition of $\M{X}$. 
\item Perform a Cramer-von-Mises (CM) or Kolmorov-Smirnov (KS) test for the null hypothesis
      $\wt{H}_0: \{ \xi_i \}_{i = 1}^{n-d} \overset{\text{i.i.d.}}{\sim} N(0, \sigma_*^2)$. Specifically, 
      compute 
      \begin{align*}
      \text{(CM)}: \; \; \int_{\R} \left( \wh{F}(x) - \Phi\left(\frac{x}{\sigma_*} \right) \right)^2 \; \frac{1}{\sigma_*}  \phi \left( \frac{x}{\sigma_*} \right) dx, \qquad
      \text{or} \; \text{(KS)}: \; \; \sup_{x \in \R} \left|\wh{F}(x) - \Phi\left(\frac{x}{\sigma_*} \right) \right|, 
      \end{align*}
      where $\phi$ and $\Phi$ denote the density and cumulative distribution, respectively, of the $N(0,1)$-distribution, and
      $\wh{F}$ denotes the empirical cumulative distribution function of the $\{ \xi_i \}_{i = 1}^{n-d}$.  
\end{enumerate}
For model \eqref{eq:limo_perm} with Gaussian design, i.e., 
\begin{equation}\label{eq:limo_perm_gauss}
y_i = \M{x}_{\pi^*(i)}^{\T} \beta^* + \sigma_* \eps_i, \quad \M{x}_i \sim N(0,I_d), \; \; i=1,\ldots,n,
\end{equation}
results in \cite{Pananjady2016} imply that the power of the KS tends to one under the alternative
hypothesis as $n \rightarrow \infty$ and $\nnorm{\beta^*}_2 - \sigma_*  \sqrt{\log(n-d)} \rightarrow \infty$. Note that 
at least for small $k$, this condition appears inevitable since $\max_{1 \leq i \leq n-d} |\zeta_i|/\sigma_* = O_{\p}\left(\sqrt{\log(n-d)} \right)$.
\begin{prop}\label{prop:testing}
Suppose model \eqref{eq:limo_perm_gauss} holds. Then for any $ t \in (0,k)$, we have 
\begin{equation*}
    \nnorm{\M{U}^{\T} \Pi^* \M{X} \beta^*}_2^2 \leq  \frac{n-d}{n} \cdot\frac{t}{2} \nnorm{\beta^*}_2^2 
\end{equation*}
with probability at most
\begin{equation*}
6 \exp\left(-\frac{k}{10}  \left[\log \frac{k}{t} + \frac{t}{k} - 1 \right] \right) + \exp(-(n-d)/24).
\end{equation*}
\end{prop}
\noindent Observe that the above proposition implies a high probability lower bound on the quantity $\nnorm{\M{U}^{\T} \Pi^* \M{X} \beta^*}_{\infty}$ appearing in \eqref{eq:derivation_test}, which
affirms the claim preceding the proposition. Empirical results depicted in Figure \ref{fig:power_gof} corroborate the importance of the ratio $\nnorm{\beta^*}_2 / \sigma_*$ for the power of the proposed test. If the latter is too small, the power hardly increases with $n$ for a fixed fraction
of mismatches $k/n$. 

\section{EM Algorithm}
The pseudo-likelihood \eqref{eq:likelihood} corresponds to the ``regular" likelihood of a mixture model, and inherits 
the computational properties of the latter. In particular, likelihood maximization in mixture models is non-convex, and thus one
cannot hope to find the global optimum of \eqref{eq:likelihood} in practice. At the same time, established computational approaches for fitting mixture models like the EM algorithm can be employed for finding approximate maximizers of \eqref{eq:likelihood} in practice. The usual convergence properties of the EM algorithm continue to hold regardless of the fact that \eqref{eq:likelihood} is a mis-specified likelihood. In fact, for the derivation of the E-step we can simply pretend that 
the $\{ (\M{x}_i, y_i)\}_{i = 1}^n$ are independent observations following the mixture distributions \eqref{eq:GMM}: the key property of the EM algorithm to increase the likelihood at each iteration does not require the likelihood to be correctly specified. In the following paragraphs, we provide the specifics of the proposed computational scheme. We first note
that given the indicator variables $\{ z_i \}_{i = 1}^n$, the complete data negative (pseudo) log-likelihood in 
$(\beta, \sigma^2, \alpha)$ is given by 
\begin{align}\label{eq:completedata_likelihood}
\begin{split}
 \su \bigg\{ &(1-z_i) \Big( -\log(1-\alpha) + \frac{\log(\sigma^2) }{2}  + \frac{(y_i - \M{x}_i^{\T} \beta)^2}{2\sigma^2} \Big) +  \\
 \qquad \qquad &z_i \Big(-\log(\alpha) + \frac{\log(\sigma^2 + \nnorm{\beta}_2^2)}{2}  + \frac{y_i^2}{2 (\sigma^2 + \nnorm{\beta}_2^2)} \Big) \bigg\}.       
\end{split} 
\end{align}

\subsection{Both $\sigma_*^2$ and $\nnorm{\beta^*}_2$ known}\label{subsec:allknown}
Denote $\tau_*^2 = \sigma_*^2 + \nnorm{\beta^*}_2^2$. Given current iterates $\wh{\beta}^{(k)}, \wh{\alpha}^{(k)}$, the E-step
is given by 
\begin{equation}\label{eq:Estep_allknown}
\pi_i^{(k)} \coloneq \E^{(k)}[z_i] =    \frac{  \displaystyle\frac{\wh{\alpha}^{(k)}}{\tau_*}   \exp\left(-\frac{y_i^2}{2 \tau_*^2} \right)}{\displaystyle \frac{\wh{\alpha}^{(k)} }{\tau_*}  \exp\left(-\frac{y_i^2}{2 \tau_*^2} \right) + \frac{(1 - \wh{\alpha}^{(k)})}{\sigma_*} \exp\left(-\frac{(y_i - \M{x}_i^{\T} \wh{\beta}^{(k)}) ^2}{2 \sigma_*^2} \right)}, \;\;\; i=1,\ldots,n,  
\end{equation}
where $\E^{(k)}$ denotes the expectation if the unknown parameters of the underlying distribution were given by 
$(\wh{\alpha}^{(k)}, \wh{\beta}^{(k)})$. Accordingly, in light of \eqref{eq:completedata_likelihood} the M-step is given by the optimization problem 
\begin{equation}\label{eq:Mstep_allknown}
\min_{\alpha}    \su \left( -(1 - \pi_i^{(k)}) \log(1 - \alpha) -  \pi_i^{(k)} \log(\alpha) \right) + \min_{\beta} \su (1 - \pi_i^{(k)}) (y_i - \M{x}_i^{\T} \beta)^2.  
\end{equation}
Both optimization problems have a closed form solution. The optimization problem in $\beta$ amounts to a weighted least
squares fit of (predictors, response)-pairs $(\M{x}_i, y_i)_{i = 1}^n$ and weights $\omega_i^{(k)} = 1 - \pi_i^{(k)}$, $i=1,\ldots,n$. This yields the updates 
\begin{equation*}
\wh{\alpha}^{(k+1)} \leftarrow \frac{1}{n} \su \pi_i^{(k)}, \qquad \wh{\beta}^{(k+1)} \leftarrow (\M{X}^{\T}  \M{W}^{(k)}  \M{X})^{-1} \M{X}^{\T} \M{W}^{(k)} \M{y}, \quad \M{W}^{(k)} = \text{diag}(\omega_1^{(k)}, \ldots, \omega_n^{(k)}),     
\end{equation*}
which is well in line with the robust regression viewpoint in $\S$\ref{subsec:robust}. Alternatively, the M-step
\eqref{eq:Mstep_allknown} can be performed subject to the additional constraint $\nnorm{\beta}_2^2 \leq \nnorm{\beta^*}_2^2$. The latter is straightforward to accommodate.  

\subsection{Plug-in approach}\label{subsec:plugin}
It is straightforward to 
estimate $\tau_*^2$ via 
\begin{equation*}
\wh{\tau}^2 = \frac{1}{n} \su y_i^2
\end{equation*}
since $\E[y_i^2] = \tau_*^2, \; i=1,\ldots,n$. After substituting $\tau_*^2$ by the above estimator and $\sigma_*^2$ by an iterate $\wh{\sigma}^{2(k)}$, the scheme of the previous subsection can still be applied. The counterpart to the E-step \eqref{eq:Estep_allknown} is given by  
\begin{equation}\label{eq:Estep_plugin}
\pi_i^{(k)} \leftarrow    \frac{  \displaystyle\frac{\wh{\alpha}^{(k)}}{\wh{\tau}}   \exp\left(-\frac{y_i^2}{2 \wh{\tau}^2} \right)}{\displaystyle \frac{\wh{\alpha}^{(k)}}{\wh{\tau}}  \exp\left(-\frac{y_i^2}{2 \wh{\tau}^2} \right) + \frac{(1 - \wh{\alpha}^{(k)})}{\wh{\sigma}^{(k)}} \exp\left(-\frac{(y_i - \M{x}_i^{\T} \wh{\beta}^{(k)}) ^2}{2 \wh{\sigma}^{2(k)}} \right)}, \;\;\; i=1,\ldots,n,      
\end{equation}
and the iterate for $\sigma^2$ is updated as 
\begin{equation*}
\wh{\sigma}^{2(k+1)} \leftarrow \frac{1}{ \su (1 - \pi_i^{(k)})} \su (1 - \pi_i^{(k)}) (y_i - \M{x}_i^{\T} \wh{\beta}^{(k)})^2. 
\end{equation*}

\subsection{Simultaneous estimation of all parameters}\label{subsec:simultaneous}
The plug-in approach of the previous section is convenient due to its closed form
updates by means of a reduction to weighted least squares estimation. In addition, it 
does not require assumptions on the distribution of the $\{ \M{x}_i \}_{i = 1}^n$. However, for isotropic Gaussian design, the plug-in approach essentially disregards the part of the complete data log-likelihood
that is associated with the $\{ z_i \}_{i = 1}^n$. It is a cleaner, but also computationally significantly more involved approach to avoid the use of the auxiliary (and eventually redundant) parameter $\tau_*^2$, and to integrate all terms of
the complete data log-likelihood \eqref{eq:completedata_likelihood}. While the E-step
\eqref{eq:Estep_plugin} remains unchanged with the only modification that 
$\wh{\tau}^2$ gets replaced by $\nnorm{\wh{\beta}^{(k)}}_2^2 + \wh{\sigma}^{2(k)}$, the M-step has no longer a closed form solution for $\wh{\beta}^{(k+1)}, \sigma^{(k+1)}$. Instead, the latter result as the minimizers of the optimization problem
\vskip.5ex
{\small \begin{equation}\label{eq:completedata_nodecoupling}
\min_{\substack{\beta \in \R^d \\ \sigma^2 > 0}}   
 \su (1-\pi_i^{(k)}) \frac{\log(\sigma^2) }{2}  + \su (1 - \pi_i^{(k)})\frac{(y_i - \M{x}_i^{\T} \beta)^2}{2\sigma^2} +  \su \pi_i^{(k)}  \frac{\log(\sigma^2 + \nnorm{\beta}_2^2)}{2}  + \su \pi_i^{(k)} \frac{y_i^2}{2 (\sigma^2 + \nnorm{\beta}_2^2)}.       
\end{equation}}
\vskip.5ex
\noindent This optimization problem fails to be convex. Rather than solving this problem,
we suggest to update the parameters via one iteration of Fisher scoring, which is also known as Titterington's algorithm in the literature on the EM algorithm \cite{Titterington1984}. Specifically, we consider the following update: 
\begin{align}\label{eq:fisher_scoring_step}
\left[\begin{array}{c}
\wh{\beta}^{(k+1)} \\[1ex]
\wh{\sigma}^{(k+1)}
\end{array} \right] = \left[\begin{array}{c}
\wh{\beta}^{(k)} \\[1ex]
\wh{\sigma}^{(k)}
\end{array} \right] + \gamma^{(k)} d^{(k)}, \qquad  d^{(k)} \coloneq -F^{(k)} \, g^{(k)},
\end{align}
where the step-size $\gamma^{(k)} \in [0,1]$ is chosen by backtracking line search (e.g., $\S$1.2 in \cite{Bertsekas1999}), and the update direction $d^{(k)}$ depends on the gradient $g^{(k)}$ of the expected complete data negative log-likelihood as well as $F^{(k)}$, the expected Fisher information (with respect to the current parameters $\wh{\beta}^{(k)}$ and  $\wh{\sigma}^{(k)}$). Since the gradient of the \emph{expected complete} data log-likelihood is known to coincide with the gradient of the \emph{incomplete} data log-likelihood (\cite{Lange1995}, p.~426), the above update scheme reduces the latter at each iteration. Expressions for $g^{(k)}$ and $F^{(k)}$ are provided in Appendix \ref{app:gradhess}.  
\subsection{Initialization}\label{subsec:init}
While the EM iterations above can be shown to yield descent at each iteration, they are not guaranteed to produce the global minimizer of the incomplete data negative log-likelihood \eqref{eq:logpseudolikelihood}. As a result, careful initialization, i.e., 
choice of the initial iterates $\wh{\beta}^{(0)}$, $\wh{\sigma}^{2(0)}$, and $\wh{\alpha}^{(0)}$ can greatly benefit the performance. As a starting point, 
one might consider  $\wh{\beta}^{(0)} = \wh{\beta}^{\text{LS}}$, where 
$\wh{\beta}^{\text{LS}} = (\M{X}^{\T} \M{X})^{-1} \M{X}^{\T} \M{y}$ denotes the ordinary least squares estimator, i.e., the naive approach that ignores the presence of mismatches. The result below indicates that under a uniform-at-random model for $\Pi^*$, this naive approach is still useful
to the extent that $\frac{\wh{\beta}^{\text{LS}}}{\nnorm{\wh{\beta}^{\text{LS}}}_2}$ provides an essentially unbiased estimator of $\frac{\beta^{*}}{\nnorm{\beta^*}_2}$. 
\begin{prop}\label{prop:naive} Consider model \eqref{eq:limo_perm_gauss} and suppose that $\Pi^*$ is chosen uniformly at random according to assumption (A1), and let $\wh{\beta}^{\text{\emph{LS}}} = (\M{X}^{\T} \M{X})^{-1} \M{X}^{\T} \M{y}$ denote the ordinary least squares estimator. We then have
\begin{equation*}
\E_{\M{X}, \eps, \pi^*}[\wh{\beta}^{\text{\emph{LS}}}] = (1-\alpha_*) \beta^*, \qquad
\cov_{\M{X}, \eps, \pi^*}[\wh{\beta}^{\text{\emph{LS}}}] =  \frac{c_*^2}{n-d} I_d + O(\nnorm{\beta^*}_2^2 / n^2), 
\end{equation*}
where $c_*^2 = (2\alpha_* - \alpha_*^2) \nnorm{\beta^*}_2^2 + \sigma_*^2$. 
\end{prop}
\noindent Proposition \ref{prop:naive} suggests $\wh{\beta} = \frac{1}{1 - \alpha_*} \wh{\beta}^{\text{LS}}$ as an unbiased estimator. Since $\alpha_*$ is typically unknown and generally not easy to estimate, an alternative is 
\begin{equation}\label{eq:betahat_rescaled}
\wh{\beta} = \frac{\wh{\beta}^{\text{LS}}}{\nnorm{\wh{\beta}^{\text{LS}}}_2} \cdot \wh{\nnorm{\beta^*}_2}, \qquad \wh{\nnorm{\beta^*}_2} = \left( \frac{1}{n} \su y_i^2 - \sigma_*^2 \right)^{1/2}, 
\end{equation}
which requires knowledge of $\sigma_*^2$ (if $\nnorm{\beta^*}_2^2 \gg \sigma_*^2$, the variance of the errors $\sigma_*^2$ can be disregarded). While potentially giving 
rise to an unbiased estimator, Proposition \ref{prop:naive} also asserts that the variance of the components of $\wh{\beta}^{\text{LS}}$ (and in turn the MSE) is rather
substantial, growing with $\nnorm{\beta^*}_2^2$ and $\alpha_*$. In particular, this implies that $\wh{\beta}^{\text{LS}}$ and its re-scaled counterparts discussed above exhibit a poor statistical efficiency relative to the oracle estimators based on knowledge of $\Pi^*$ or the set of correct matches $\{1 \leq i \leq n:\, z_i = 0 \}$. In light of this, another option is to employ robust regression methods like Huber's estimator as considered in \cite{SlawskiBenDavid2017} even though the latter is limited to the regime
of small to moderate $\alpha_*$.

\noindent \emph{Connection to the Lahiri-Larsen estimator.} In their seminal work on regression with linked data \cite{Lahiri05}, Lahiri and Larsen propose the following estimator 
\begin{equation}\label{eq:LL}
\wh{\beta}^{\text{LL}} = (\M{X}^{\T} \M{Q}^{\T} \M{Q} \M{X})^{-1} \M{X}^{\T} \M{Q}^{\T} \M{y}, \quad \text{where} \; \M{Q} = \E[\Pi^*].    
\end{equation}
It is easy to see that the above estimator is unbiased, i.e.,  
$\E_{\pi^*, \eps}[\wh{\beta}^{\text{LL}}] = \beta^*$ uniformly in $\M{X}$. Assuming that $\Pi^*$ is drawn uniformly at random from the set of $k$-sparse permutations of $n$ elements, the matrix $\M{Q}$ is given by
\begin{equation*}
\M{Q} = \left(1 - \alpha_* - \frac{\alpha_*}{n-1} \right) I_n + \frac{\alpha_*}{n-1} \M{1} \M{1}^{\T}.    
\end{equation*}
Discarding all terms in $\M{Q}$ involving $\alpha_*/(n-1)$, the estimator
\eqref{eq:LL} reduces to the estimator in Proposition \ref{prop:naive}. It is not hard to establish asymptotic equivalence of the two estimators; the formal derivation is omitted for the sake of brevity. 

\begin{figure}
\begin{center}
\includegraphics[height = 0.24\textheight]{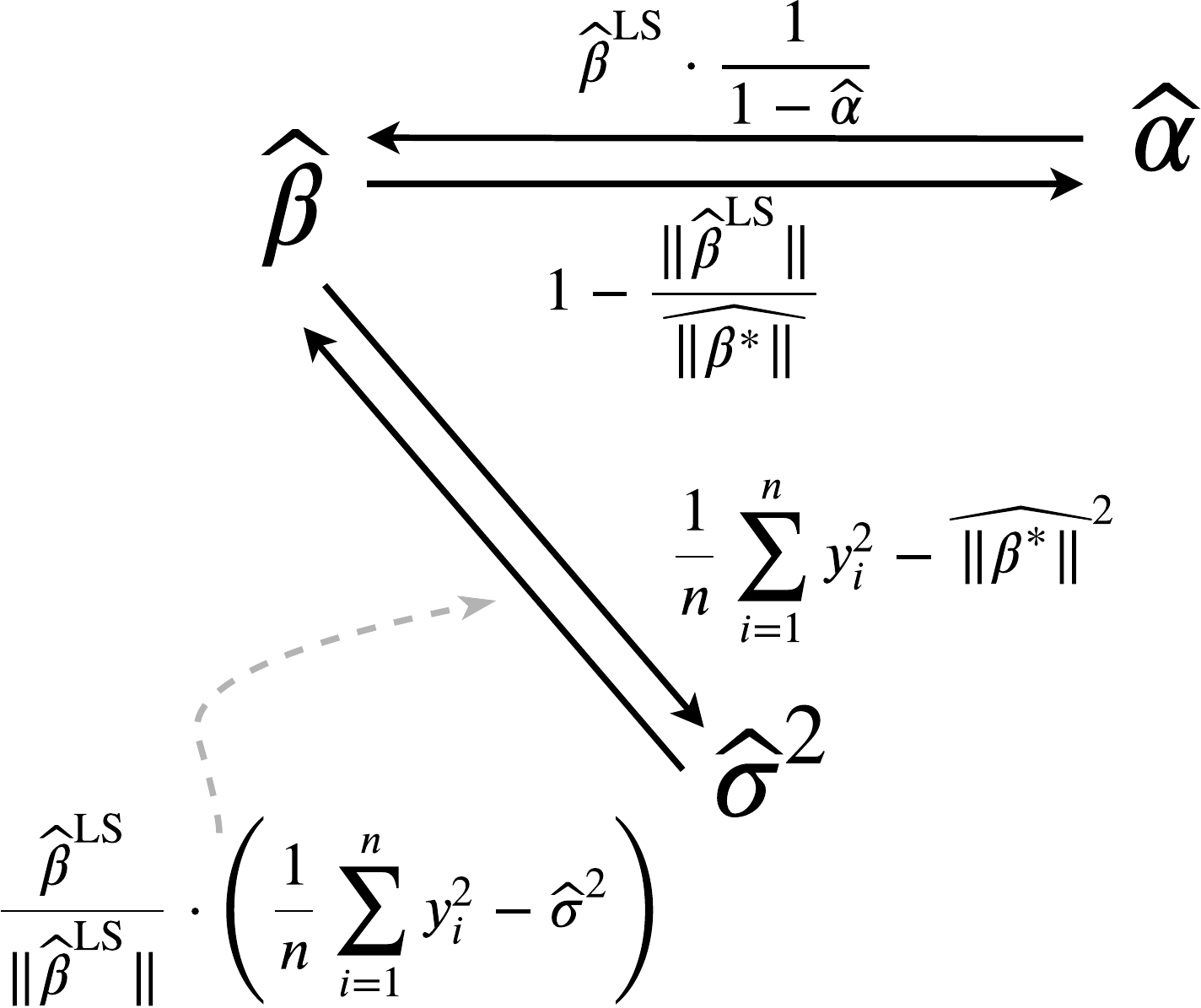}
\end{center}
\vspace*{-4ex}
\caption{Diagram visualizing the interdependence for initial estimators 
$(\wh{\beta}, \wh{\sigma}, \wh{\alpha})$ in light of Proposition \ref{prop:naive} and relation \eqref{eq:betahat_rescaled}. Note that given an estimator $\wh{\beta}$, one can use $\wh{\nnorm{\beta^*}_2} = \nnorm{\wh{\beta}}_2$.}\label{fig:init}
\end{figure}

Equipped with an estimator of $\beta^*$, the quantities $\alpha_*$ and 
$\sigma_*$ can be estimated according to Figure \ref{fig:init}. Estimation 
of the three quantities is generally interdependent in the sense that
one of the three parameters is supposed to be known or accurately estimable. The latter requirement becomes significantly more difficult to meet as the fraction of mismatches $\alpha_*$ increases. 

\section{Empirical results}\label{sec:verify}

\subsection{Gaussian design}\label{subsec:synthetic}
Data is generated according to model \eqref{eq:limo_perm_gauss} with $n = 200$, $d = 10$, and $\beta^*$ drawn uniformly at
random from the corresponding sphere. We vary $\sigma_* \in \{.01,.1,.2,.5,1\}$ and $\alpha_* \in \{.1,.2,\ldots,.7\}$, and 
the permutation $\pi^*$ is drawn uniformly at random according to (A1). For each configuration of $(\sigma_*, \alpha_*)$, 100 independent replications are considered. In order
to estimate the parameters $(\beta_*, \sigma_*, \alpha_*)$, we consider the approaches in $\S$\ref{subsec:plugin} and 
$\S$\ref{subsec:simultaneous}. The former is computationally simpler as it reduces to solving a sequence of weighted least
squares problems. Both approaches were initialized with $\wh{\beta}^{(0)} = \wh{\beta}^{\text{LS}}$, $\wh{\sigma}^{(0)} = 
n^{-1/2} \nnorm{\M{y} - \M{X} \wh{\beta}^{(0)}}_2$, and $\wh{\alpha}^{(0)} = 0.5$. More sophisticated initialization schemes as discussed in $\S$\ref{subsec:init} did not yield substantial gains in performance. The approaches in $\S$\ref{subsec:plugin} and $\S$\ref{subsec:simultaneous} are compared to the oracle estimator 
\begin{equation}\label{eq:oracle_estimator}
\wh{\beta}^{\text{o}} = \left( \M{X}^{\T} \M{X} \right)^{-1} \M{X}^{\T} \Pi^{*\T} \M{y}, 
\end{equation}
and the 
robust regression method in the recent paper \cite{SlawskiBenDavid2017}; the latter is given an additional advantage by
equipping it with knowledge of the noise level $\sigma_*$, which is required for the optimal choice of the regularization parameter \cite{SlawskiBenDavid2017}. 
\vskip2ex
\noindent \emph{Results}. Tables \ref{tab:errors_beta} and \ref{tab:errors_sigmaalpha} display the $\ell_2$-estimation 
errors $\nnorm{\wh{\beta} - \beta^*}_2 / \nnorm{\wh{\beta}^{\text{o}} - \beta^*}_2$ relative to that of the oracle estimator as well as the relative error for the noise level $|\wh{\sigma}/\sigma_* - 1|$ and absolute error for the fraction of mismatches $|\wh{\alpha} - \alpha_*|$. The tables contain medians of these measures of error over 100 independent replications. Bootstrap standard errors of the medians are given in parentheses as superscript, and the fraction of outliers among the 100 replications are given as subscript; here, outliers are defined as observations exceeding the 3rd quartile by more than 1.5 times the interquartile range, a criterion that is commonly used to depict outliers in boxplots \cite{Tukey1977}. Table \ref{tab:errors_beta} shows that for both variants of the proposed approach, the
estimation error for the regression coefficients in $\ell_2$-norm are largely within factors of three or less of
the oracle estimator. The errors can be seen to vary more strongly with the fraction of mismatches $\alpha_*$ than
with the noise level $\sigma_*$. For example, for $\alpha_* = 0.1$, the errors are within a factor of 1.3 of the oracle,
and increase to a factor of 3 and higher as $\alpha_*$ reaches $0.6$. Table \ref{tab:errors_beta} also indicates that 
the computationally more complex approach in $\S$\ref{subsec:simultaneous} performs better than the plug-in approach in 
 $\S$\ref{subsec:plugin}, with visible differences for the smallest value of $\sigma_*$ under consideration 
 ($\sigma_* = 0.01$). As $\sigma_*$ increases, these differences vanish. Both approaches significantly outperform the robust regression method \cite{SlawskiBenDavid2017} whose performance degrades much more severely with the fraction of mismatches $\alpha_*$. The differences are most pronounced for $\sigma_* \in \{0.01, 0.1, 0.2\}$, and partially disappear for
 $\sigma_* = 1$. The latter observation can be explained by the fact that in this setting, $\nnorm{\beta^*}_2 / \sigma_* = 1$ in which case the error induced by mismatches is of the same order as that induced by additive noise. Table \ref{tab:errors_sigmaalpha} shows that the proposed approach also enables estimation of the parameters $\sigma_*$ and
 $\alpha_*$ with small error in most settings. For $\alpha_* \in \{0.6,0.7\}$ and/or $\sigma_* = 1$, estimation becomes
 a serious challenge and as result, the estimators $\wh{\alpha}$ and $\wh{\sigma}$ become less reliable. 
\begin{table}
\begin{minipage}{\textwidth}
    \centering
    (i) $\nnorm{\wh{\beta} - \beta^*}_2 / \nnorm{\wh{\beta}^{\text{o}} - \beta^*}_2$, approach in $\S$\ref{subsec:simultaneous} \\[.6ex]
    \begin{tabular}{|c||l|l|l|l|l|l|l|}
    \hline
    \mbox{{\Large $\bsfrac{\alpha_*}{\sigma_*}$}}     &  $0.1$ & $0.2$ & $0.3$ & $0.4$ & $0.5$ & $0.6$ & $0.7$ \\
    \hline
     0.01  & $1.05_{(.02)}^{(.01)}$    & $1.5_{(.03)}^{(.03)}$  & $1.26_{(0)}^{(.03)}$  &  $1.28_{(.02)}^{(.04)}$
     & $1.39_{(.03)}^{(0.03)}$& $1.91_{(.04)}^{(0.05)}$ & $2.39_{(.08)}^{(.1)}$   \\
     \hline
     0.1   & $1.08_{(.01)}^{(.03)}$  & $1.21_{(.03)}^{(.02)}$  & $1.34_{(.03)}^{(.03)}$  & $1.45_{(.03)}^{(.04)}$ & $1.58_{(.05)}^{(.05)}$  & $2.26_{(.02)}^{(.09)}$   & $3.25_{(.09)}^{(.18)}$ \\
     \hline
     0.2   & $1.11_{(.03)}^{(.02)}$  & $1.24_{(.03)}^{(.03)}$  & $1.38_{(.03)}^{(.04)}$  & $1.56_{(.01)}^{(.07)}$ & $1.72_{(.02)}^{(.06)}$  &  $2.38_{(.08)}^{(.11)}$  & $3.08_{(.13)}^{(.17)}$ \\
     \hline
     0.5   & $1.15_{(.05)}^{(.02)}$ & $1.29_{(.03)}^{(.03)}$  & $1.42_{(.06)}^{(.04)}$  & $1.73_{(.02)}^{(.05)}$ &  $1.99_{(.07)}^{(.06)}$ & $2.88_{(.03)}^{(.16)}$   & $4.69_{(.06)}^{(.24)}$ \\
     \hline
     1     & $1.24_{(.03)}^{(.03)}$  & $1.38_{(.02)}^{(.04)}$  & $1.55_{(.04)}^{(.04)}$  & $1.82_{(.06)}^{(.09)}$  &  $2.03_{(.06)}^{(.1)}$ &  $2.80_{(.06)}^{(.11)}$   & $4.00_{(.01)}^{(.22)}$  \\
     \hline
    \end{tabular}
    \vskip2ex
    (ii) $\nnorm{\wh{\beta} - \beta^*}_2 / \nnorm{\wh{\beta}^{\text{o}} - \beta^*}_2$, approach in $\S$\ref{subsec:plugin} \\[.6ex]
    \begin{tabular}{|c||l|l|l|l|l|l|l|}
    \hline
    \mbox{{\Large $\bsfrac{\alpha_*}{\sigma_*}$}}     &  $0.1$ & $0.2$ & $0.3$ & $0.4$ & $0.5$ & $0.6$ & $0.7$ \\
    \hline
     0.01  & $1.27_{(.06)}^{(1.2)}$    & $1.69_{(0)}^{(1.2)}$  & $1.83_{(.02)}^{(2.3)}$  &  $2.60_{(.03)}^{(2.0)}$
     & $1.89_{(.04)}^{(0.57)}$& $3.13_{(.06)}^{(1.2)}$ & $4.12_{(.09)}^{(2.4)}$   \\
     \hline
     0.1   & $1.19_{(.11)}^{(.04)}$  & $1.31_{(.12)}^{(.08)}$  & $1.53_{(.05)}^{(.06)}$  & $1.7_{(.08)}^{(.06)}$ & $1.8_{(.04)}^{(.1)}$  & $2.53_{(.09)}^{(.11)}$   & $3.48_{(.11)}^{(.16)}$ \\
     \hline
     0.2   & $1.13_{(.05)}^{(.02)}$  & $1.3_{(.04)}^{(.04)}$  & $1.44_{(.06)}^{(.05)}$  & $1.59_{(.03)}^{(.07)}$ & $1.74_{(.02)}^{(.07)}$  &  $2.49_{(.04)}^{(.14)}$  & $3.88_{(.15)}^{(.26)}$ \\
     \hline
     0.5   & $1.17_{(.05)}^{(.02)}$ & $1.3_{(.03)}^{(.03)}$ & $1.41_{(.06)}^{(.04)}$  & $1.76_{(.03)}^{(.05)}$ & $2.02_{(.08)}^{(.07)}$  & $2.72_{(.03)}^{(.18)}$ &  $4.63_{(.03)}^{(.17)}$ \\
     \hline
     1     & $1.24_{(.03)}^{(.04)}$  & $1.38_{(.03)}^{(.03)}$  & $1.55_{(.06)}^{(.05)}$  & $1.87_{(.03)}^{(.1)}$  &  $2.02_{(.06)}^{(.11)}$ &  $2.84_{(.06)}^{(.13)}$   & $4.07_{(0)}^{(.26)}$  \\
     \hline
    \end{tabular}
    \vskip2ex
    (iii) $\nnorm{\wh{\beta}^{H} - \beta^*}_2 / \nnorm{\wh{\beta}^{\text{o}} - \beta^*}_2$, approach in \cite{SlawskiBenDavid2017} \\[.6ex]
    \begin{tabular}{|c||l|l|l|l|l|l|l|}
    \hline
    \mbox{{\Large $\bsfrac{\alpha_*}{\sigma_*}$}}     &  $0.1$ & $0.2$ & $0.3$ & $0.4$ & $0.5$ & $0.6$ & $0.7$ \\
    \hline
     0.01  & $2.4_{(.02)}^{(.07)}$    & $3.99_{(.02)}^{(.15)}$  & $6.9_{(.04)}^{(.23)}$  &  $10.4_{(.04)}^{(.4)}$
     & $21.4_{(.07)}^{(1.3)}$& $132_{(.02)}^{(6.6)}$ & $272_{(.01)}^{(12)}$   \\
     \hline
     0.1   & $2.2_{(.05)}^{(.05)}$  & $3.64_{(.02)}^{(.12)}$  & $6.0_{(.02)}^{(.25)}$  & $8.6_{(.02)}^{(.3)}$ & $12.7_{(.06)}^{(.35)}$  & $19.2_{(.04)}^{(.69)}$   & $28.9_{(.01)}^{(1.2)}$ \\
     \hline
     0.2   & $2.0_{(.05)}^{(.06)}$  & $3.1_{(.01)}^{(.13)}$  & $5.1_{(.02)}^{(.17)}$  & $6.8_{(.02)}^{(.2)}$ & $9.2_{(.04)}^{(.25)}$  &  $12.3_{(.04)}^{(.46)}$  & $15.5_{(.02)}^{(.71)}$ \\
     \hline
     0.5   & $1.48_{(.02)}^{(.06)}$ & $2.1_{(.01)}^{(.09)}$ & $3.05_{(.03)}^{(.08)}$  & $3.8_{(.04)}^{(.11)}$ & $4.67_{(.04)}^{(.13)}$  & $5.73_{(0)}^{(.25)}$ &  $6.85_{(.02)}^{(.3)}$ \\
     \hline
     1     & $1.14_{(.01)}^{(.04)}$  & $1.38_{(.01)}^{(.06)}$  & $1.81_{(.03)}^{(.05)}$  & $2.14_{(.03)}^{(.07)}$  &  $2.53_{(.04)}^{(.08)}$ & $3.01_{(0)}^{(.13)}$   & $3.69_{(.01)}^{(.15)}$  \\
     \hline
    \end{tabular}    
    \caption{Median estimation errors for $\beta^*$ in $\ell_2$-norm relative to the estimation error of the oracle estimator $\wh{\beta}^{\text{o}}$ \eqref{eq:oracle_estimator} in dependence of $\alpha_*$ (columns) and $\sigma_*$ (rows). Bootstrap standard errors are given in superscripts, fraction of outliers in subscript (cf.~the paragraph ``Results" in $\S$\ref{subsec:synthetic} for a definition).}
    \label{tab:errors_beta}
\end{minipage}
\end{table}
\begin{table}
    \centering
    ($\sigma_*$) $|\wh{\sigma}/\sigma_* - 1|$, approach in $\S$\ref{subsec:simultaneous} \\[.6ex]
    \begin{tabular}{|c||l|l|l|l|l|l|l|}
    \hline
    \mbox{{\Large $\bsfrac{\alpha_*}{\sigma_*}$}}     &  $0.1$ & $0.2$ & $0.3$ & $0.4$ & $0.5$ & $0.6$ & $0.7$ \\
    \hline
     0.01  & $.04_{(.04)}^{(.00)}$    & $.05_{(0)}^{(.01)}$  & $.04_{(0)}^{(.01)}$  &  $.06_{(.01)}^{(.01)}$
     & $.06_{(0)}^{(.01)}$& $.08_{(.01)}^{(.01)}$ & $.12_{(.08)}^{(.01)}$   \\
     \hline
     0.1   & $.04_{(.01)}^{(.01)}$  & $.06_{(0)}^{(.01)}$  & $.05_{(0)}^{(.01)}$  & $.08_{(.02)}^{(.01)}$ & $.09_{(.02)}^{(.01)}$  & $0.1_{(0)}^{(.01)}$   & $.23_{(.08)}^{(.04)}$ \\
     \hline
     0.2   & $.04_{(.03)}^{(.00)}$  & $.06_{(0)}^{(.01)}$  & $.06_{(.01)}^{(.01)}$  & $.07_{(.06)}^{(.01)}$ & $0.1_{(.04)}^{(.01)}$  &  $.14_{(.02)}^{(.01)}$  & $.26_{(.06)}^{(.03)}$ \\
     \hline
     0.5   & $.05_{(.01)}^{(.01)}$ & $.07_{(0)}^{(.01)}$  & $.07_{(.02)}^{(.01)}$  & $.09_{(.02)}^{(.01)}$ &  $.11_{(.06)}^{(.01)}$ & $.21_{(.01)}^{(.03)}$   & $.43_{(0)}^{(.04)}$ \\
     \hline
     1     & $.07_{(.01)}^{(.01)}$  & $.11_{(0)}^{(.01)}$  & $.08_{(.01)}^{(.01)}$  & $.11_{(.08)}^{(.01)}$  &  $0.1_{(.09)}^{(.01)}$ &  $.14_{(.1)}^{(.02)}$   & $0.2_{(.11)}^{(.01)}$  \\
     \hline
    \end{tabular}
    \vskip2ex
($\alpha_*$) $|\wh{\alpha} - \alpha_*|$, approach in $\S$\ref{subsec:simultaneous}\\[.6ex]
    \begin{tabular}{|c||l|l|l|l|l|l|l|}
    \hline
    \mbox{{\Large $\bsfrac{\alpha_*}{\sigma_*}$}}     &  $0.1$ & $0.2$ & $0.3$ & $0.4$ & $0.5$ & $0.6$ & $0.7$ \\
    \hline
     0.01  & $.00_{(.04)}^{(.00)}$    & $.01_{(.01)}^{(.00)}$  & $.01_{(.02)}^{(.00)}$  &  $.01_{(.09)}^{(.00)}$
     & $.01_{(.02)}^{(.00)}$& $.01_{(.02)}^{(.00)}$ & $.01_{(.07)}^{(.00)}$   \\
     \hline
     0.1   & $.01_{(.01)}^{(.00)}$  & $.01_{(.03)}^{(.00)}$  & $.01_{(0.02)}^{(.00)}$  & $.02_{(.04)}^{(.00)}$ & $.02_{(.01)}^{(.00)}$  & $.02_{(0)}^{(.00)}$   & $.03_{(.07)}^{(.00)}$ \\
     \hline
     0.2   & $.01_{(.03)}^{(.00)}$  & $.02_{(.02)}^{(.00)}$  & $.02_{(.02)}^{(.00)}$  & $.02_{(.01)}^{(.00)}$ & $.03_{(.01)}^{(.00)}$  &  $.03_{(.01)}^{(.00)}$  & $.04_{(.06)}^{(.01)}$ \\
     \hline
     0.5   & $.02_{(.02)}^{(.00)}$ & $.03_{(.02)}^{(.00)}$  & $.03_{(.04)}^{(.00)}$  & $.04_{(.01)}^{(.00)}$ &  $.05_{(.01)}^{(.01)}$ & $.07_{(0)}^{(.01)}$   & $0.1_{(0)}^{(.01)}$ \\
     \hline
     1     & $.07_{(.02)}^{(.01)}$  & $.08_{(0)}^{(.01)}$  & $.07_{(.06)}^{(.01)}$  & $.05_{(.04)}^{(.01)}$  &  $.08_{(.11)}^{(.01)}$ &  $.15_{(.02)}^{(.01)}$   & $.22_{(0)}^{(.01)}$  \\
     \hline
    \end{tabular}
    \caption{Median relative estimation errors for $\sigma_*$ and median estimation error for $\alpha_*$ based on the approach in $\S$\ref{subsec:simultaneous}. The structure of the table follows that of the previous table (Table \ref{tab:errors_beta}). A standard error of ``.00" refers to a standard error less than $.005$.}
    \label{tab:errors_sigmaalpha}
\end{table}

\subsection{CPS wage data}
\begin{figure}
    \centering
    \begin{tabular}{cc}
    CPS & El Nino  \\
    \includegraphics[width = .35\textwidth]{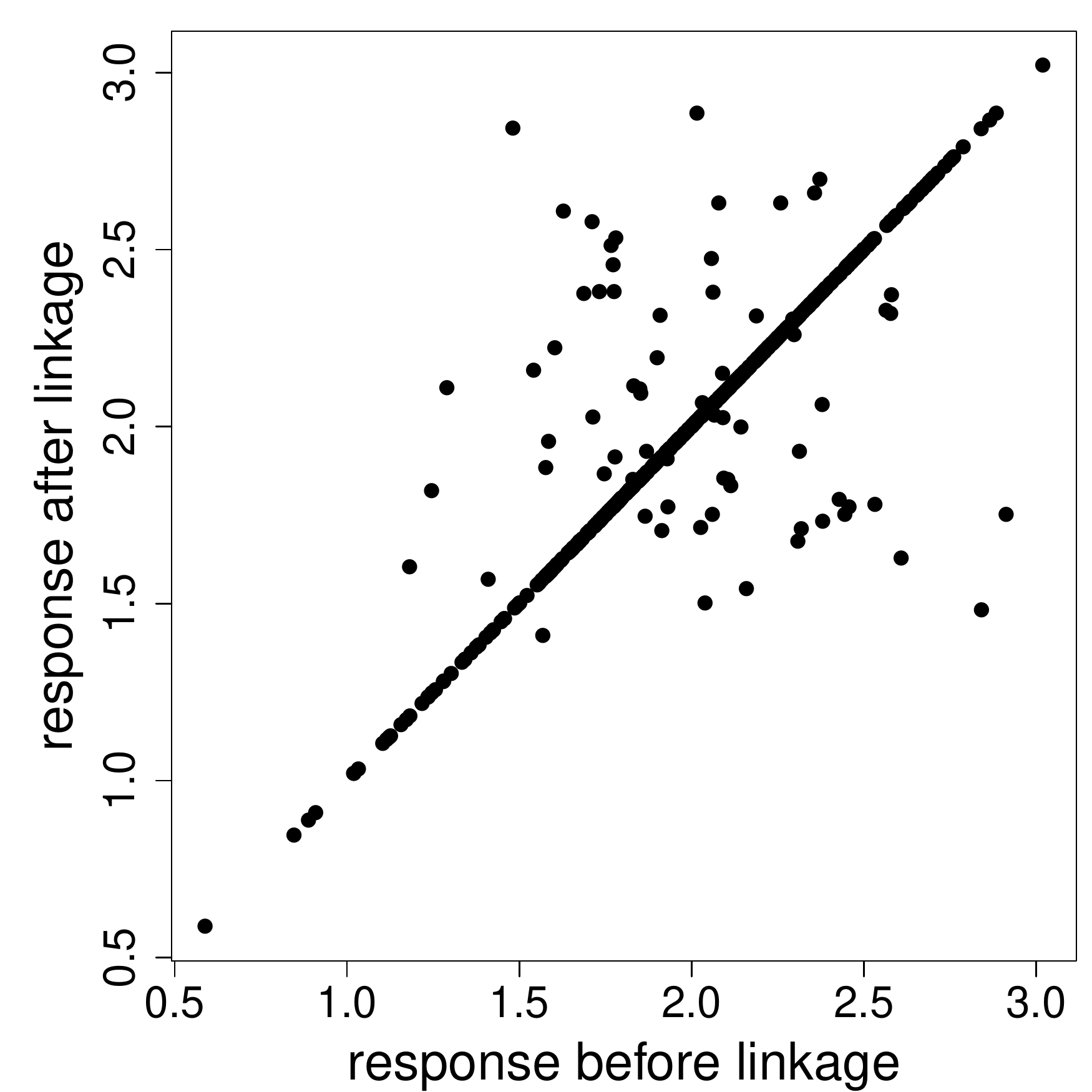} \hspace*{3ex} & \hspace*{3ex} \includegraphics[width = .35\textwidth]{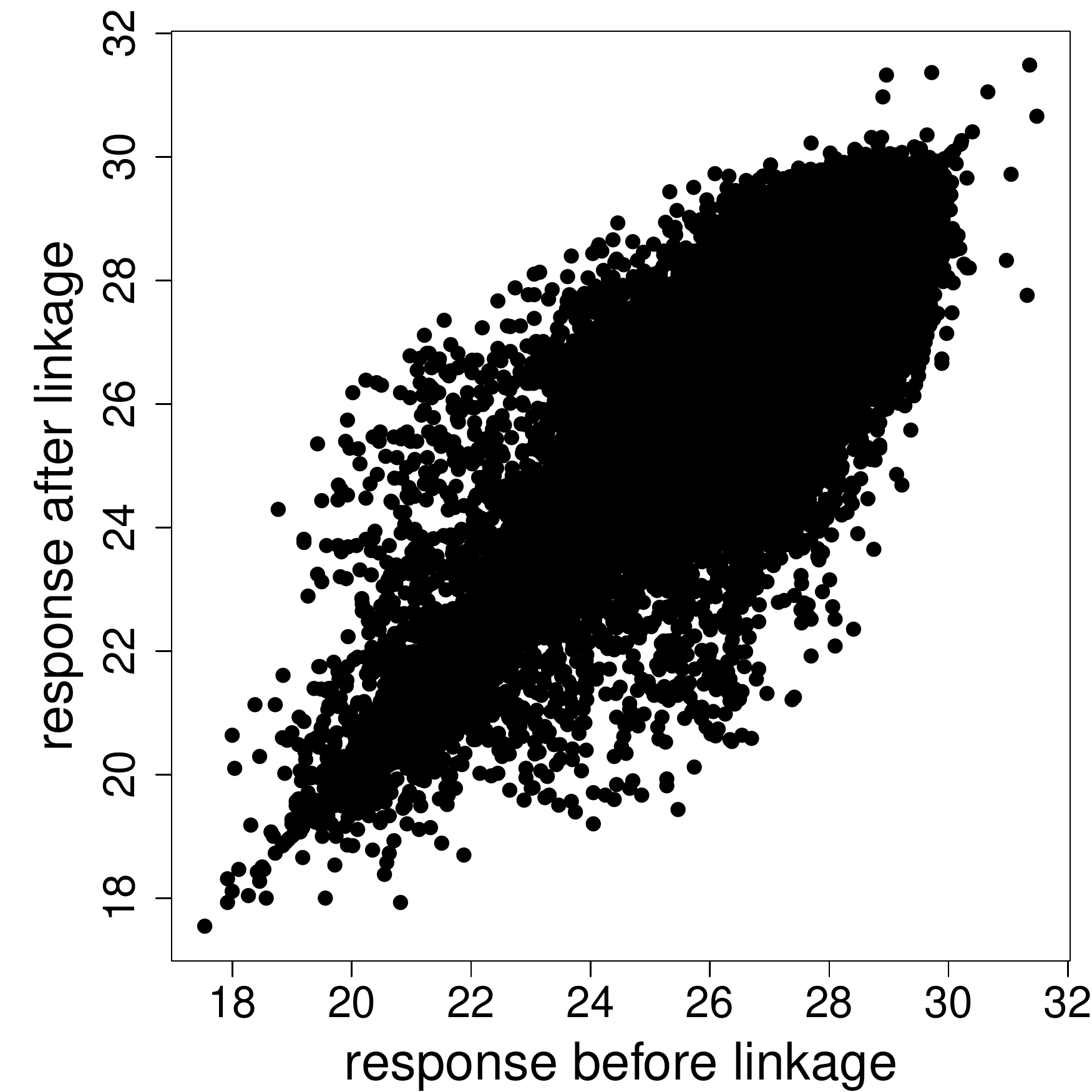}
    \end{tabular}
    \caption{Response variable before and after linkage for the CPS wage data (left) and the El Nino data (right). The angle bisector corresponds to the situation without mismatch.}
    \label{fig:response_linkage_error}
\end{figure}

We use the CPS wage data set available from STATLIB (http://lib.stat.cmu.edu/datasets/) containing information on wages 
and other characteristics of $n = 534$ workers, including sex, number of years of education, years of work experience, type of occupation and
union membership. To mimic the situation in record linkage, we complement this data set with synthetic demographic information (first name, last name, zip code etc.) generated by the R package \texttt{generator} \cite{Generator}  matching the information on sex and age in the original data set. We 
re-create the response variable $\log(\texttt{wage})$ (logarithm of the hourly wage) according to 
{\small \begin{align*}
\log(\texttt{wage}) &= \beta_0^* + \beta_1^* \cdot I(\texttt{sex}=``F") + \beta_2^* \cdot \texttt{experience} + \beta_3^* \cdot \texttt{experience}^2 + \beta_4^* \cdot\texttt{education} + \\
\qquad &+ \beta_5^* \cdot I(\texttt{occupation}=\text{{``Sales"}}) + \beta_6^* \cdot I(\texttt{occupation}=\text{``Clerical"}) \\
\qquad&+  \beta_7^* \cdot I(\texttt{occupation}=\text{``Service"}) + \beta_8^* \cdot I(\texttt{occupation}=\text{``Professional"}) \\
\qquad&+ \beta_{9}^* \cdot I(\texttt{occupation}=\text{``Other"})
+ \beta_{10}^*  \cdot I(\texttt{union}=``Y") + \sigma_* \eps, \quad \eps \sim N(0,1).
\end{align*}}
Here, $I(\ldots)$ represents indicator variables: ``F" is short for female, ``union = Y[es]" indicates membership in a union,
and the variable \texttt{occupation} represents one of six occupational categories (reference category is ``Management"). The variables
\texttt{experience} and \texttt{education} represent work experience and education (in \#years), respectively, and are both treated as numerical variables. 
The regression coefficients $\beta_0^*,\ldots,\beta_{10}^*$ were chosen as the coefficients from the least
squares fit of the same model with the original wages. By re-creating the response, we fully maintain the correlation
structure of the predictors while achieving a better model fit (the choice $\sigma_* = 1.5$ leads to an $R^2$ close to 0.7), 
which helps to demonstrate the impact of linkage error and the ability of the proposed approach to provide remedy. Linkage error is generated by splitting the entire file into two files, one of which only contains the response variable and the zip code
of the individuals while the second file contains all variables except for the response. The thus obtained two files were 
linked based on the variable zip code using the R package \texttt{fastLink} \cite{fastlink}. Since zip code does not represent a unique identifier, a fraction of $\alpha_* \approx .13$ of the records are incorrectly matched. Figure \ref{fig:response_linkage_error} displays the discrepancy between the response before and after file linkage. We compare the following approaches (i) oracle least squares
based on the original undivided file, (ii) naive least squares ignoring linkage error, (iii) the robust regression method in \cite{SlawskiBenDavid2017}, (iv) the Lahiri-Larsen (LL) estimator \eqref{eq:LL}, where the matrix $\M{Q}$ is constructed by 
assuming that matching among observations with the same zip code is done uniformly at random, and (v) the proposed approach in the variant of $\S$\ref{subsec:plugin} in which the solution of (iii) along with a robust estimator of the noise level (properly re-scaled median absolute deviation of the residuals) is used for initialization. These five approaches are compared in terms of 
$\nnorm{\wh{\beta} - \wh{\beta}^{\text{o}}}_2$ and $\su (y_i - x_{\pi^*(i)}^{\T} \wh{\beta})^2$ (mean
squared error on the original data). For a more detailed comparison, regression coefficients and standard errors of (i), (ii) and (iii) are reported in Table \ref{tab:cps}.
\begin{table}[]
    \centering
    \begin{tabular}{|c||c|c|c|c|c|}        
    \hline
    & oracle & proposed & LL & robust \cite{SlawskiBenDavid2017} & naive \\ 
    \hline
    $\nnorm{\wh{\beta} - \wh{\beta}^{\text{o}}}_2$  & 0  & 0.03 & 0.07 &    0.17 & 0.20 \\
    \hline
    $\su (y_i - x_{\pi^*(i)}^{\T} \wh{\beta})^2$    &  23.46 & 23.57 & 23.68 & 24.22    &  24.82 \\
    \hline
    \end{tabular}
    \vskip3ex
    \begin{tabular}{|c|c|c|c|c|c|c|c|c|c|c|c|c|c|}
    \hline
                & $\wh{\beta}_0$  & $\wh{\beta}_1$ & $\wh{\beta}_2$ & $\wh{\beta}_3$ & $\wh{\beta}_4$ &  $\wh{\beta}_5$ & $\wh{\beta}_6$ & $\wh{\beta}_7$ &  $\wh{\beta}_8$ & $\wh{\beta}_9$ & $\wh{\beta}_{10}$ & $\wh{\sigma}^2$ & $\wh{\alpha}$ \\
                \hline  & & & & & & & & & & & & & \\[-3ex]
    {\small oracle}          & $\overset{\mbox{{\normalsize 1.01}}}{\mbox{{\scriptsize 8.3e-2}}}$ & $\sss{-.21}{2e-2}$   & $\sss{.03}{2.6e-3}$ & $\sss{-.0004}{5.8e-5}$&  $\sss{.07}{4.9e-3}$ & $\sss{-.32}{4.5e-2}$  & $\sss{-.23}{3.8e-2}$   & $\sss{-.37}{4e-2}$  & $\sss{-.05}{3.6e-2}$  & $\sss{-.20}{3.7e-2}$ & $\sss{.20}{2.5e-2}$ & $\sss{.045}{3e-3}$  & NA\\
    \hline
    & & & & & & & & & & & & & \\[-3ex]
    {\small proposed}          & $\sss{1.02}{9.8e-2}$ & $\sss{-.22}{2.4e-2}$ & $\sss{.03}{3.5e-3}$  & $\sss{-.0005}{8e-5}$   & $\sss{.07}{6e-3}$  &   $\sss{-.30}{6.3e-2}$ & $\sss{-.23}{3.5e-2}$   & $\sss{-.38}{4.6e-2}$ & $\sss{-.06}{3.1e-2}$ & $\sss{-.19}{3.4e-2}$ & $\sss{.21}{2.3e-2}$  &$\sss{.043}{4.4e-3}$  & $\sss{.14}{2.7e-2}$\\
    \hline
    & & & & & & & & & & & & & \\[-3ex]
    {\small naive}             &  $\sss{1.2}{1e-1}$  & $\sss{-.19}{2.6e-2}$  & $\sss{-.03}{3.3e-3}$  &  $\sss{-.0004}{7.3e-5}$  & $\sss{.06}{6.2e-3}$ & $\sss{-.26}{5.8e-2}$  & $\sss{-.23}{4.8e-2}$    & $\sss{-.35}{5.1e-2}$ & $\sss{-.08}{4.5e-2}$ & $\sss{-.18}{4.8e-2}$ & $\sss{.18}{3.1-2}$ & $\sss{.072}{4e-3}$  & NA \\
    \hline 
    \end{tabular}
    \caption{Summary of results for the CPS wage data. Top: $\ell_2$-estimation error for the regression coefficients and mean squared error. Bottom: Parameter estimates and their standard errors (small font size) of the proposed method
    in comparison to oracle and naive least squares.}
    \label{tab:cps}
\end{table}

\noindent \emph{Results}. The figures in Table \ref{tab:cps} show that the proposed approach exhibits similar
performance as the oracle estimator. The estimated regression coefficients and their standard errors are rather
close, and the fraction of mismatches is also estimated accurately ($\wh{\alpha} = .14$ compared to $\alpha_* = .13$).
For comparison, the changes in the regression coefficients are more noticeable for the naive least squares solution,
which also yields a considerably reduced fit as is indicated by an inflation of the estimated residual variance 
($0.072$ compared to $0.045$). The robust regression method in \cite{SlawskiBenDavid2017} yields improvements
relative to the naive approach, but they are less substantial relative to the proposed approach. The latter also
outperforms the Lahiri-Larsen estimator, which is equipped with additional information in terms of the matrix $\M{Q}$.  

\subsection{El Nino data}
We here build on the case study presented in $\S$3.2 in \cite{SlawskiBenDavid2017} that is based on the El Nino data set \cite{Dua2017}. The latter contains meteorological measurements recorded by a sensor network known as the Tropical Atmosphere Ocean Array consisting of $\sim$70 buoys placed across the equatorial Pacific. Sensors positioned at those buoys record
zonal and meridional wind speeds (abbreviated \texttt{zon} and \texttt{mer}), relative humidity (\texttt{humidity}), air temperature (\texttt{air.temp}), sea surface temperature and subsurface temperatures down to a depth of 500 meters (\texttt{s.s.temp}). The regression model considered in \cite{SlawskiBenDavid2017} is given by 
\begin{equation*}
\texttt{air.temp} = \beta_0^* + \beta_{\texttt{z}}^* \cdot \texttt{zon.winds} + \beta_{\texttt{m}}^* \cdot \texttt{mer.winds} + \beta_{\texttt{h}}^* \cdot \texttt{humidity} + \beta_{\texttt{s}}^* \cdot \texttt{s.s.temp} + \eps.    
\end{equation*}
Each set of measurements is uniquely identified by the buoy identifier and the day of its recording. In \cite{SlawskiBenDavid2017} this information is discarded, and the response variable (\texttt{air.temp}) is 
put into a separate file that additionally contains the longitude and latitude of the measurement as an inexact 
identifier. The latter is used subsequently as matching variable by \texttt{fastLink} \cite{fastlink} to merge the response
variable with the predictor variables. The right panel of Figure \ref{fig:response_linkage_error} shows that the 
error induced by mismatches is rather substantial, with a fraction of .82 of the $n = 93,935$ observations being mismatched. However, not all mismatches lead to substantial changes in the response: for example, only a fraction of .16 of the observations is associated with an error in the response larger than twice the residual standard error from the oracle least squares fit. The fact that the majority of mismatches does not lead to major errors is ultimately a consequence of the fact that meteorological
measurements sharing the same (latitude, longitude)-pair exhibit spatial correlation even though they may not correspond to the same observational unit. 
\begin{table}[]
    \centering
    \begin{tabular}{|c||c|c|c|c|c|}        
    \hline
    & oracle & proposed & LL & robust \cite{SlawskiBenDavid2017} & naive \\ 
    \hline
    $\nnorm{\wh{\beta} - \wh{\beta}^{\text{o}}}_2$  & 0  & 0.04 & 1.56 & 0.59 & 1.57 \\
    \hline
    $\su (y_i - x_{\pi^*(i)}^{\T} \wh{\beta})^2$    &  24.4e3 & 24.7e3  & 25.2e3 & 25.1e3    &  25.9e3 \\
    \hline
    \end{tabular}
    \vskip3ex
    \begin{tabular}{|c|c|c|c|c|c|c|c|}
    \hline
                & $\wh{\beta}_0$  & $\wh{\beta}_{\texttt{z}}$ & $\wh{\beta}_{\texttt{m}}$ & $\wh{\beta}_{\texttt{h}}$ & $\wh{\beta}_{\texttt{s}}$ &  $\wh{\sigma}^2$ & $\wh{\alpha}$ \\
                \hline  & & & & & & &  \\[-3ex]
    {\small oracle}         & $\sss{5.15}{4.9e-2}$ &$\sss{-.056}{5.5e-4}$ &$\sss{-.031}{5.8e-4}$ & $\sss{-.022}{3.4e-4}$ & $\sss{.844}{1.1e-3}$ & $\sss{.259}{1.2e-3}$ & NA \\
    \hline
    & & & & & & &  \\[-3ex]
    {\small proposed}    & $\sss{5.12}{7.3e-2}$ & $\sss{-.044}{8.1e-4}$ & $\sss{-.038}{7.8e-4}$ & $\sss{-.016}{4.8e-4}$ & $\sss{.827}{1.7e-3}$ & $\sss{.358}{2.6e-3}$ &  $\sss{.073}{1.4e-3}$     \\
    \hline
    & & & & & & &  \\[-3ex]
    {\small naive}     & $\sss{6.72}{7.5e-2}$ & $\sss{-.037}{8.3e-4}$ & $\sss{-.045}{8.9e-4}$ & $\sss{-.017}{5.1e-4}$& $\sss{.774}{1.7e-3}$   &  $\sss{.594}{2.7e-3}$   & NA  \\
    \hline 
    \end{tabular}
    \caption{  Summary of results for the El Nino data. Top: $\ell_2$-estimation error for the regression coefficients and mean squared error. Bottom: Parameter estimates and their standard errors (small font size) of the proposed method
    in comparison to oracle and naive least squares.}
    \label{tab:elnino}
\end{table}

\noindent \emph{Results.} According to Table \ref{tab:elnino}, the proposed method is not far from the oracle
estimator. The estimates for the regression coefficients are noticeably closer than those of the naive approach.
The latter yields a poor fit, with the residual variance inflated by more than a factor of two ($0.594$ vs.~$0.259$). 
The proposed approach also outperforms the method in \cite{SlawskiBenDavid2017} as well as the Lahiri-Larsen estimator
(with $\M{Q}$ constructed analogously to the previous subsection, with zip code replaced by (longitude, latitude)) in terms of the $\ell_2$-estimation error and mean squared error. The performance of the Lahiri-Larsen is suboptimal here due to 
a small number of distinct (latitude, longitude)-pairs relative to the sample size (about 5k vs.~94k); by construction
of the matrix $\M{Q}$, the Lahiri-Larsen estimator here amounts to averaging predictors and response with the same
(latitude, longitude), and a subsequent weighted least squares fit with the thus obtained averages. As a result, the
effective sample size is reduced to 5k. Lastly, it is worth noting that the proposed approach estimates the fraction
of mismatches as approximately $.073$, whereas the underlying fraction of mismatches is around $.82$. This gap is
a consequence of the fact that the majority of mismatches do not substantially change the response compared to
the error of the regression model as explained above. The estimate of $.073$ turns out to be close to the fraction
of mismatches that change the response by three times the residual standard error or more. 
\section{Conclusion}\label{sec:conc}
In this paper, we have presented a pseudo-likelihood method to account for mismatches in the response variables in linear regression, an important problem in the analysis of linked files. 
The proposed method is computationally scalable, requires at most minimum tuning, provides estimators of all parameters of interest, and achieves promising empirical performance according to the results in the preceding section. In light of these appealing properties, we hope that the method will be widely adopted to deal with the scenarios discussed herein. Owing to its simple modular structure, the method considered herein can be generalized to a variety of other regression models including multiple response variables, generalized linear models and non-parametric regression, which will be investigated in future work. Another interesting direction of research concerns the adjustment for mismatches in the situation where a subset of the predictor variables is contained in the same file as the response, while the remaining predictors are contained in a separate file.    








%
\bibliographystyle{plain}
\bibliography{references_M}

\begin{thebibliography}{10}

\bibitem{Abid2017}
A.~Abid, A.~Poon, and J.~Zou.
\newblock {Linear regression with shuffled labels}.
\newblock arXiv:1705.01342, 2017.

\bibitem{Abid2018}
A.~Abid and J.~Zou.
\newblock {Stochastic EM for shuffled linear regression}.
\newblock In {\em Allerton Conference on Communication, Control, and
  Computing}, pages 470--477, 2018.

\bibitem{Balakrishnan1962}
A.~Balakhrisnan.
\newblock On the problem of time jitter in sampling.
\newblock {\em IRE Transactions on Information Theory}, 8:226--236, 1962.

\bibitem{Bertsekas1999}
D.~Bertsekas.
\newblock {\em Nonlinear Programming}.
\newblock Athena Scientific, 2nd edition edition, 1999.

\bibitem{Burkard2009}
R.~Burkard, M.~Dell'Amico, and S.~Martello.
\newblock {\em Assignment Problems: Revised Reprint}.
\newblock SIAM, 2009.

\bibitem{Carpentier2016}
A.~Carpentier and T.~Schl\"uter.
\newblock Learning relationships between data obtained independently.
\newblock In {\em Proceedings of the International Conference on Artifical
  Intelligence and Statistics (AISTATS)}, pages 658--666, 2016.

\bibitem{Chambers2009}
R.~Chambers.
\newblock Regression analysis of probability-linked data.
\newblock Technical report, Statistics New Zealand, 2009.

\bibitem{Chen2009}
J.~Chen and P.~Li.
\newblock {Hypothesis test for normal mixture models: The EM approach}.
\newblock {\em The Annals of Statistics}, 37:2523--2542, 2009.

\bibitem{DasGupta2003}
S.~DasGupta and A.~Gupta.
\newblock {An elementary proof of a theorem of Johnson and Lindenstrauss}.
\newblock {\em Random Structures and Algorithms}, 22:60--65, 2003.

\bibitem{DeGroot1971}
M.~DeGroot, P.~Feder, and P.~Goel.
\newblock Matchmaking.
\newblock {\em The Annals of Mathematical Statistics}, 42:578--593, 1971.

\bibitem{DeGroot1976}
M.~DeGroot and P.~Goel.
\newblock {The matching problem for multivariate normal data}.
\newblock {\em Sankhya, Series B}, 38:14--29, 1976.

\bibitem{DeGroot1980}
M.~DeGroot and P.~Goel.
\newblock Estimation of the correlation coefficient from a broken random
  sample.
\newblock {\em The Annals of Statistics}, 8:264--278, 1980.

\bibitem{Dua2017}
D.~Dheeru and E.~Karra Taniskidou.
\newblock {UCI} machine learning repository, 2017.

\bibitem{Emiya2014}
V.~Emiya, A.~Bonnefoy, L.~Daudet, and R.~Gribonval.
\newblock {Compressed sensing with unknown sensor permutation}.
\newblock In {\em Acoustics, Speech and Signal Processing (ICASSP)}, pages
  1040--1044, 2014.

\bibitem{fastlink}
E.~Enamorado, B.~Eifield, and K.~Imai.
\newblock {Fast Probabilistic Record Linkage with Missing Data}.
\newblock R-package, Version 0.2.0.

\bibitem{Goel1975}
P.~Goel.
\newblock {On re-pairing observations in a broken sample}.
\newblock {\em The Annals of Statistics}, 3:1364--1369, 1975.

\bibitem{Goel1987}
P.~Goel and T.~Ramalingam.
\newblock Some properties of the maximum likelihood strategy for re-pairing a
  broken random sample.
\newblock {\em {Journal of Statistical Planning and Inference}}, 16:237--248,
  1987.

\bibitem{Haghighatshoar2017}
S.~Haghighatshoar and G.~Caire.
\newblock {Signal Recovery from Unlabeled Samples}.
\newblock In {\em International Symposium on Information Theory (ISIT)}, pages
  451--455, 2017.

\bibitem{Generator}
Paul Hendricks.
\newblock {\em {\texttt{generator}: Generate Data Containing Fake Personally
  Identifiable Information}}, 2015.
\newblock R package version 0.1.0.

\bibitem{Hof2014}
M.~Hof and A.~Zwinderman.
\newblock {A mixture model for the analysis of data derived from record
  linkage}.
\newblock {\em Statistics in Medicine}, 34:74--92, 2015.

\bibitem{Hof12}
M.~H.~P. Hof and A.~H. Zwinderman.
\newblock Methods for analyzing data from probabilistic linkage strategies
  based on partially identifying variables.
\newblock {\em Statistics in Medicine}, 31(30):4231--4242, 2012.

\bibitem{Hsu2017}
D.~Hsu, K.~Shi, and X.~Sun.
\newblock Linear regression without correspondence.
\newblock In {\em Advances in Neural Information Processing Systems (NIPS)},
  pages 1531--1540, 2017.

\bibitem{Lahiri05}
P.~Lahiri and Michael~D. Larsen.
\newblock Regression analysis with linked data.
\newblock {\em Journal of the American Statistical Association},
  100(469):222--230, 2005.

\bibitem{Lange1995}
K.~Lange.
\newblock {A gradient algorithm locally equivalent to the EM algorithm}.
\newblock {\em Journal of the Royal Statistical Society Series B}, 57:425--437,
  1995.

\bibitem{Lindsay1988}
B.~Lindsay.
\newblock {Composite likelihood methods}.
\newblock {\em Contemporary Mathematics}, 80:221--239, 1988.

\bibitem{Yohai2006}
R.~Maronna, R.~Martin, and V.~Yohai.
\newblock {\em {Robust Statistics: Theory and Methods}}.
\newblock Wiley, 2006.

\bibitem{Neter65}
J.~Neter, S.~Maynes, and R.~Ramanathan.
\newblock The effect of mismatching on the measurement of response error.
\newblock {\em Journal of the American Statistical Association}, 60:1005--1027,
  1965.

\bibitem{Pananjady2017}
A.~Pananjady, M.~Wainwright, and T.~Cortade.
\newblock {Denoising linear models with permuted data}.
\newblock arXiv:1704.07461, 2017.

\bibitem{Pananjady2016}
A.~Pananjady, M.~Wainwright, and T.~Cortade.
\newblock {Linear regression with shuffled data: Statistical and computational
  limits of permutation recovery}.
\newblock {\em IEEE Transactions on Information Theory}, 3826--3300, 2018.

\bibitem{Weed2018}
P.~Rigollet and J.~Weed.
\newblock {Uncoupled isotonic regression via minimum Wasserstein
  deconvolution}.
\newblock arXiv:1806.10648; to appear in Information \& Inference, 2019.

\bibitem{Scheuren93}
F.~Scheuren and W.~Winkler.
\newblock {Regression analysis of data files that are computer matched I}.
\newblock {\em Survey Methodology}, 19:39--58, 1993.

\bibitem{Scheuren97}
F.~Scheuren and W.~Winkler.
\newblock {Regression analysis of data files that are computer matched II}.
\newblock {\em Survey Methodology}, 23:157--165, 12 1997.

\bibitem{Shi2018}
X.~Shi, X.~Lu, and T.~Cai.
\newblock Spherical regresion under mismatch corruption with application to
  automated knowledge translation.
\newblock arXiv:1810.05679, 2018.

\bibitem{SlawskiBenDavid2017}
M.~Slawski and E.~Ben-David.
\newblock {Linear regression with sparsely permuted data}.
\newblock {\em Electronic Journal of Statistics}, 13:1--36, 2019.

\bibitem{SlawskiBenDavidLi2019}
M.~Slawski, E.~Ben-David, and P.~Li.
\newblock {A Two-Stage Approach to Multivariate Linear Regression with Sparsely
  Mismatched Data}.
\newblock arXiv:1907.07148, July 2019.

\bibitem{SlawskiRahmaniLi2018}
M.~Slawski, M.~Rahmani, and Ping Li.
\newblock {A Sparse Representation-Based Approach to Linear Regression with
  Partially Shuffled Labels}.
\newblock In {\em Proceedings of the Thirty-Fifth Conference on Uncertainty in
  Artificial Intelligence (UAI)}, 2019.

\bibitem{Titterington1984}
M.~Titterington.
\newblock Recursive parameter estimation using incomplete data.
\newblock {\em Journal of the Royal Statistical Society Series B}, 46:257--267,
  1984.

\bibitem{Tsakiris2018b}
M.~Tsakiris.
\newblock Eigenspace conditions for homomorphic sensing.
\newblock arXiv:1812.07966, December 2018.

\bibitem{Tukey1977}
J.~Tukey.
\newblock {\em {Exploratory Data Analysis}}.
\newblock Addison Wesley, 3rd edition, 1977.

\bibitem{Unnikrishnan2015}
J.~Unnikrishnan, S.~Haghighatshoar, and M.~Vetterli.
\newblock Unlabeled sensing with random linear measurements.
\newblock {\em IEEE Transactions on Information Theory}, 64:3237--3253, 2018.

\bibitem{Unnikrishnan2013}
J.~Unnikrishnan and M.~Vetterli.
\newblock Sampling and reconstruction of spatial fields using mobile sensors.
\newblock {\em IEEE Transactions on Signal Processing}, 61:2328--2340, 2013.

\bibitem{vanderVaart1998}
A.~van~der Vaart.
\newblock {\em {Asymptotic Statistics}}.
\newblock Cambridge University Press, 1998.

\bibitem{Varin2011}
C.~Varin, N.~Reid, and D.~Firth.
\newblock An overview of composite likelihood estimation.
\newblock {\em Statistica Sinica}, 21:5--42, 2011.

\bibitem{Wu1998}
Y.~N. Wu.
\newblock A note on broken sample problem.
\newblock Technical report, Department of Statistics, University of Michigan,
  1998.

\bibitem{ZhangSlawskiLi2019}
H.~Zhang, M.~Slawski, and P.~Li.
\newblock {Permutation recovery from multiple measurement vectors in unlabeled
  sensing}.
\newblock In {\em IEEE International Symposium on Information Theory (ISIT)},
  2019.

\bibitem{Zhu2004}
H.-T. Zhu and H.~Zhang.
\newblock Hypothesis testing in mixture regression models.
\newblock {\em Journal of the Royal Statistical Society Series B}, 66:3--16,
  2004.

\end{thebibliography}

\appendix
\section{Proof of Proposition \ref{prop:testing}}
Let us recall that $\M{U}$ denotes the $n$-by-$(n-d)$ matrix whose columns 
form an orthonormal basis of the orthogonal complement of the column space of $\M{X}$.
Let further $\M{Q}$ be a $d$-by-$d$ orthonormal matrix such that $\M{Q}^{\T} \beta^* = e_1$, 
where $e_1 = (1,0,\ldots,0)^{\T}$ denotes the first canonical basis vector.
Following the proof of Lemma 5 in \cite{Pananjady2016}, we have 
\begin{align*}
\nnorm{\M{U}^{\T} \Pi^* \M{X} \beta^*}_2^2 \,\overset{\mc{D}}{=} 
\nnorm{\texttt{P}_{\M{X}}^{\perp} \Pi^* \M{X} \beta^*}_2^2 
= \nnorm{\texttt{P}_{\Pi^* \M{X} \M{Q}}^{\perp} \, \M{X} \M{Q} \M{Q}^{\T} \beta^*}_2^2  = \nnorm{\texttt{P}_{\Pi^* \wt{\M{X}}}^{\perp} \wt{\M{X}} e_1}_2^2 \, \nnorm{\beta^*}_2^2,   
\end{align*}
where $\wt{\M{X}} = \M{X} \M{Q} \overset{\mc{D}} = \M{X}$ by the rotational invariance of the Gaussian distribution, with $\overset{\mc{D}}{=}$ denoting
equality in distribution. In the sequel, we will lower bound the term $\nnorm{\texttt{P}_{\Pi^* \wt{\M{X}}}^{\perp} \wt{\M{X}} e_1}_2^2$. For this purpose, we define 
\begin{equation*}
\M{z} = \wt{\M{X}}_{:,1}, \qquad S_1 = \text{range}(\Pi^* \wt{\M{X}} e_1), \qquad S_{-1} = \text{range}(\Pi^* \wt{\M{X}}_{:,2:d}), \qquad
S_{-1}^{\perp} = \text{range}(\Pi^* \wt{\M{X}}_{:,2:d})^{\perp}, 
\end{equation*}
where $:,1$ and $:,2:d$ refer to the column submatrices formed from column one and columns two to $d$, respectively. With this notation, we have
\begin{align}
  \nnorm{\texttt{P}_{\Pi^* \wt{\M{X}}}^{\perp} \wt{\M{X}} e_1}_2^2 = \nnorm{\texttt{P}_{\Pi^* \wt{\M{X}}}^{\perp} \M{z}}_2^2 \, \nnorm{\beta^*}_2^2
                                               =  \nnorm{\texttt{P}_{S_1^{\perp} \cap S_{-1}^{\perp}} \M{z}}_2^2  \, \nnorm{\beta^*}_2^2 
                                               &=  \nnorm{\texttt{P}_{S_1^{\perp} \cap S_{-1}^{\perp}} \texttt{P}_{S_1^\perp} \M{z}}_2^2 \, \nnorm{\beta^*}_2^2 \notag \\
                                               &=   \nnorm{\texttt{P}_{S_1^{\perp} \cap S_{-1}^{\perp}} \M{u}}_2^2 \cdot \nnorm{\texttt{P}_{S_1^\perp} \M{z}}_2^2 \cdot \nnorm{\beta^*}_2^2 \label{eq:prop_testing_decomp},
\end{align}
where $\M{u} = \texttt{P}_{S_1^\perp} \M{z} / \nnorm{P_{S_1^\perp} \M{z}}_2$. By Lemma 3 in \cite{Pananjady2016}, we have the following for the middle term in  
\eqref{eq:prop_testing_decomp}: 
\begin{equation*}
\p(\nnorm{\texttt{P}_{S_1^\perp} \M{z}}_2^2 \leq t) \leq  6 \exp\left(-\frac{k}{10}  \left[\log \frac{k}{t} + \frac{t}{k} - 1 \right] \right) \; \; \forall t \geq 0.    
\end{equation*}
It remains to provide a lower bound on the first term in \eqref{eq:prop_testing_decomp}. We note
that conditional on $\M{z}$, $S_{1}^{\perp}$ becomes a fixed subspace, and
$S_1^{\perp} \cap S_{-1}^{\perp}$ becomes a subspace of dimension $n-d$ chosen uniformly at random within the subspace $S_1^{\perp}$, whose dimension is $n-1$. According to well-known
results on random projections \cite{DasGupta2003}, we have for any $\epsilon \in (0,1)$, conditional on $\M{z}$
\begin{equation*}
 \p\left((1-\epsilon) \frac{n-d}{n-1} \leq \nnorm{\texttt{P}_{S_1^{\perp} \cap S_{-1}^{\perp}} \M{u}}_2^2 \leq (1 + \epsilon) \frac{n-d}{n-1} \right) \geq 1 - \exp\left(-(n-d)(\epsilon^2/4 - \epsilon^3/6) \right).
\end{equation*}
Since the right is independent of $\M{z}$, we conclude the proof by setting $\epsilon = 1/2$. 

\section{Proof of Proposition \ref{prop:naive}}
We start by proving the expression for the expectation. Under the assumption that $\pi^*$ is chosen uniformly 
at random from the set of $k$-sparse permutations, we have 
\begin{equation}\label{eq:E_y}
  \E_{\pi^*, \bm{\epsilon}}[y_i | \mathbf{X}] = (1 - \alpha_*) \M{x}_i^{\T} \beta^* + \frac{\alpha_*}{n-1} \sum_{j \neq i} \M{x}_j^{\T} \beta^*, \quad i=1,\ldots,n.
\end{equation}
We can write this as
\begin{equation*}
  \E_{\pi^*, \bm{\epsilon}}[\mathbf{y} | \mathbf{X}] =  \left(1 - \alpha_* - \frac{\alpha_*}{n-1} \right) \mathbf{X} \beta^* + \frac{1}{n-1} \alpha_* \M{1} \M{1}^{\T} \M{X} \beta^*.
\end{equation*}
Hence
\begin{align}\label{eq:expectation_naive}
\begin{split}
  \E_{\mathbf{X}}[\E_{\pi^*, \bm{\epsilon}}[\wh{\beta}^{\text{LS}} | \mathbf{X}]] &= \E_{\mathbf{X}}[ (\mathbf{X}^{\T} \M{X})^{-1} \M{X}^{\T} \E_{\pi^*, \bm{\epsilon}}[\mathbf{y} | \mathbf{X}] ] \\
                                                                          &= \left(1 - \alpha_* - \frac{\alpha_*}{n-1} \right)  \beta^* + \alpha_* \frac{n}{n-1} \E_{\M{X}}\left[ (\M{X}^{\T} \M{X})^{-1} \M{X}^{\T} \frac{\M{1} \M{1}^{\T}}{n} \M{X} \beta^* \right] \\
                                                                          &= \left(1 - \alpha_* - \frac{\alpha_*}{n-1} \right)  \beta^* + \alpha_* \frac{n}{n-1} \frac{\beta^*}{n}  \\                                                                          &=(1 - \alpha_*) \beta^*.
\end{split}                                                                          
\end{align}
The fact that the expectation with respect to $\M{X}$ in the second line equals to $\beta^* / n$ can be seen as follows: the term inside the expectation is the regression coefficient
of a regression $(1/n) \M{x}^{\T} \beta^*$ on $\M{x}$, where $\M{x} \sim N(0, I_d)$.\\
We next derive the expression for $\cov(\wh{\beta}^{\text{LS}})$. The variance decomposition formula implies that 
\begin{align}\label{eq:variance_decomposition}
\cov(\wh{\beta}^{\text{LS}}) = \E_{\mathbf{X}}\left[\cov_{\pi^*, \bm{\epsilon}}(\wh{\beta}^{\text{LS}} | \mathbf{X}) \right]  + 
\cov_{\mathbf{X}}(\E_{\pi^*, \bm{\epsilon}}[\wh{\beta}^{\text{LS}}|\mathbf{X}]).
\end{align}
For the second term, we obtain along the lines of display \eqref{eq:expectation_naive} that 
\begin{equation*}
\cov_{\mathbf{X}}(\E_{\pi^*, \bm{\epsilon}}[\wh{\beta}^{\text{LS}}|\mathbf{X}]) = \alpha_* \frac{n}{n-1} \cov_{\mathbf{X}}\left((\M{X}^{\T} \M{X})^{-1} \M{X}^{\T} \frac{\M{1} \M{1}^{\T}}{n} \M{X} \beta^* \right).
\end{equation*}
Now observe that 
\begin{align*}
(\M{X}^{\T} \M{X})^{-1} \M{X}^{\T} \frac{\M{1} \M{1}^{\T}}{n} \M{X} \beta^*
&=\left(\frac{\M{X}^{\T} \M{X}}{n} \right)^{-1} \bar{\M{x}} \bar{\M{x}}^{\T} \beta^*, \qquad \bar{\M{x}} \coloneq \frac{\mathbf{X}^{\T} \bm{1}}{n} \\
&= O_{\p}(1) \cdot O_{\p}(n^{-1/2})  \cdot O_{\p}(n^{-1/2}) \cdot O(\nnorm{\beta^*}_2).
\end{align*}
As a result, 
\begin{equation}\label{eq:Covbeta_lowerorder}
\cov\left((\M{X}^{\T} \M{X})^{-1} \M{X}^{\T} \frac{\M{1} \M{1}^{\T}}{n} \M{X} \beta^* \right) = O(\nnorm{\beta^*}_2^2/n^2),   
\end{equation}
i.e., the second term in \eqref{eq:variance_decomposition} becomes a lower order term.\\ 
Turning to the first term in \eqref{eq:variance_decomposition}, we have
\begin{equation}\label{eq:variance_decomposition_first}
  \cov_{\pi^*, \bm{\epsilon}}(\wh{\beta}^{\text{LS}} \,| \mathbf{X}) = (\M{X}^{\T} \M{X})^{-1} \M{X}^{\T} \cov_{\pi^*, \bm{\epsilon}}(\mathbf{y} | \mathbf{X}) \, \M{X}  (\M{X}^{\T} \M{X})^{-1}.
 \end{equation}
In the following, we compute the covariance matrix in \eqref{eq:variance_decomposition_first}. For the diagonal entries, we start with  
 \begin{equation*}
   \E_{\pi^*, \bm{\epsilon}}\left[ y_i^2 | \mathbf{X} \right] = (1 - \alpha_*) (\M{x}_i^{\T} \beta^*)^2 + \alpha_* \frac{1}{n-1}  \sum_{j \neq i} (\M{x}_j^{\T} \beta^*)^2 + \sigma_*^2,
 \end{equation*}
 and thus in virtue of \eqref{eq:E_y}
 \begin{align*}
   \var(y_i | \mathbf{X}) &= ((1 - \alpha_*) - (1 - \alpha_*)^2) (\M{x}_i^{\T} \beta^*)^2 + \alpha_* \frac{1}{n-1} \sum_{j \neq i} (\M{x}_j^{\T} \beta^*)^2
  \\&- \frac{\alpha_*^2}{(n-1)^2}  \left( \sum_{j \neq i} \M{x}_j^{\T} \beta^* \right)^2 - 2 (1 - \alpha_*) \alpha_* \frac{1}{n-1} \M{x}_i^{\T} \beta^* \sum_{j \neq i} \M{x}_j^{\T} \beta^* + \sigma_*^2. 
 \end{align*}
 Taking the expectation with respect to $\M{X}$, this term becomes
 \begin{equation}\label{eq:varyi}
\var(y_i) = \alpha_* (1 - \alpha_*) \nnorm{\beta^*}_2^2 + \alpha_* \nnorm{\beta^*}_2^2 - \frac{\alpha_*^2}{n-1} \nnorm{\beta^*}_2^2 + \sigma_*^2. 
\end{equation}
Let $p_{ij}^{kl} = \p(\pi^*(i) = k, \, \pi^*(j) = l)$, $1 \leq i,j,k,l \leq n$, $i \neq j$, $k \neq l$. We then compute 
for any $i \neq j$
\begin{align*}
\E_{\pi^*, \bm{\epsilon}}[y_i y_j| \mathbf{X}] = \left(p_{ij}^{ij} + p_{ij}^{ji} \right) \M{x}_i^{\T} \beta^* \M{x}_j^{\T} \beta^*  
+ \sum_{k \notin  \{i,j\}} (p_{ij}^{ik} + p_{ij}^{ki}) \M{x}_{i}^{\T}{\beta^*} \M{x}_k^{\T} \beta^*  
&+ \sum_{k \notin  \{i,j\}} (p_{ij}^{kj} + p_{ij}^{jk}) \M{x}_{j}^{\T}{\beta^*} \M{x}_k^{\T} \beta^*  \\
&+ \sum_{k \notin \{i,j \}}  \sum_{l \notin \{i,j,k \}} p_{ij}^{kl} \M{x}_k^{\T} \beta^*  \M{x}_l^{\T} \beta^*.
\end{align*}
By elementary probability arguments, it can be shown that the above terms simplify to 
\begin{align}
&\E_{\pi^*, \bm{\epsilon}}[y_i y_j| \mathbf{X}] = \left((1 - \alpha_*)\left(1 - \frac{n}{n-1}\alpha_* \right) + \nu_{1} \right) \M{x}_i^{\T} \beta^* \M{x}_j^{\T} \beta^*  \notag\\
&\qquad \qquad \quad + \frac{(1 - \alpha_*) \frac{n}{n-1} \alpha_* + \nu_{2} }{n-2} \M{x}_i^{\T} \beta^* \sum_{k \notin \{ i,j \}} \M{x}_k^{\T} \beta^* + \frac{(1 - \alpha_*) \frac{n}{n-1} \alpha_* + \nu_{2} }{n-2} \M{x}_j^{\T} \beta^* \sum_{k \notin \{ i,j \}} \M{x}_k^{\T} \beta^* \notag\\
&\qquad \qquad \quad  +\nu_{3} \frac{1}{n-2} \frac{1}{n-3} \sum_{k \notin \{ i,j \}} \sum_{l \notin \{ i,j,k \}} \M{x}_k^{\T} \beta^*  \M{x}_l^{\T} \beta^*, \label{eq:Eyiyj_final}
\end{align}
where $0 \leq \nu_{i} \leq \alpha_*^2$, $i = 1,2,3$.  Moreover, in order to obtain 
$\cov_{\pi^*, \bm{\epsilon}}(y_i, y_j | \M{X})$, $i \neq j$, we compute
\begin{align}
\E_{\pi^*, \bm{\epsilon}}[y_i | \mathbf{X}] \cdot \E_{\pi^*, \bm{\epsilon}}[y_j | \mathbf{X}] 
&= (1-\alpha_*)^2 (\M{x}_i^{\T} \beta^*) (\M{x}_j^{\T} \beta^*) + \frac{\alpha_* (1 - \alpha_*)}{n-1} (\M{x}_j^{\T} \beta^*)^2 \notag \\
&+ \frac{\alpha_* (1 - \alpha_*)}{n-1} (\M{x}_i^{\T} \beta^*)^2 
+ \frac{\alpha_*^2}{(n-1)^2} \sum_{l \neq i} \sum_{m \neq j} (\M{x}_l^{\T} \beta^*)  (\M{x}_m^{\T} \beta^*)
 \label{eq:covyiyj_final}. 
\end{align}
Let $c_*^2 = (2\alpha_* - \alpha_*^2) \nnorm{\beta^*}_2^2 + \sigma_*^2$, i.e. the leading term in
\eqref{eq:varyi}, and
let further $\Delta = \cov(\mathbf{y}|\mathbf{X}) - c_*^2 I_d$. We then have 
\begin{align}
\cov(\wh{\beta}^{\text{LS}}|\mathbf{X}) &= c_*^2 \left( \mathbf{X}^{\T} \M{X} \right)^{-1} + 
(\mathbf{X}^{\T} \mathbf{X})^{-1} \mathbf{X}^{\T} \Delta \mathbf{X} (\mathbf{X}^{\T} \mathbf{X})^{-1} \notag\\
&=c_*^2 \left( \mathbf{X}^{\T} \M{X} \right)^{-1} + \left(\frac{\mathbf{X}^{\T} \mathbf{X}}{n} \right)^{-1} \frac{\mathbf{X}^{\T} \Delta \mathbf{X}}{n} \left(\frac{\mathbf{X}^{\T} \mathbf{X}}{n} \right)^{-1} \cdot \frac{1}{n} \notag\\
&=c_*^2 \left( \mathbf{X}^{\T} \M{X} \right)^{-1} + O_{\p}(1) \cdot O_{\p}(\nnorm{\beta^*}_2^2/n) \cdot O_{\p}(1) \cdot (1/n) \notag\\
&=c_*^2 \left( \mathbf{X}^{\T} \M{X} \right)^{-1} +  O_{\p}(\nnorm{\beta^*}_2^2/n^2) \label{eq:covbetaX},
\end{align}
where the order $O_{\p}(\nnorm{\beta^*}_2^2/n)$ in the third line follows from stitching together \eqref{eq:varyi}, \eqref{eq:Eyiyj_final}, and \eqref{eq:covyiyj_final}. When taking the expectation of \eqref{eq:covbetaX} with respect to $\M{X}$, we use that $(\M{X}^{\T} \M{X})^{-1}$ follows an inverse Wishart distribution. In combination with the
variance decomposition \eqref{eq:variance_decomposition} and result \eqref{eq:Covbeta_lowerorder}, this yields  
\begin{equation*}
\cov(\wh{\beta}^{\text{LS}}) = \frac{c_*^2}{n-d} I_d + O(\nnorm{\beta^*}_2^2 / n^2). 
\end{equation*}
\section{Expressions for computing the estimators in \eqref{eq:sandwich}}\label{app:sandwich}
We here provide expressions for the case
in which the marginal density of the response variable $f_y$ is known (cf.~also $\S$\ref{subsec:allknown}). For simplicity, we 
drop the subscript $n$ from $\wh{\theta}_n$. Moreover, for 
$\theta \in \R^d \times [0, \infty) \times [0,1]$, we write
\begin{equation*}
\pi_i(\theta) = \frac{\alpha f_y(y_i)}{\alpha f_y(y_i) + (1 -\alpha) \frac{1}{\sqrt{2 \pi}\sigma} \exp\left(-\frac{(y_i - \M{x}_i^{\T} \beta) ^2}{2 \sigma^{2}} \right)} = \frac{\alpha \wt{f}_y(y_i)}{\alpha \wt{f}_y(y_i) + (1 -\alpha) \frac{1}{\sigma} \exp\left(-\frac{(y_i - \M{x}_i^{\T} \beta) ^2}{2 \sigma^{2}} \right)},
\end{equation*}
$i=1,\ldots,n$, where $\wt{f}_y = \sqrt{2 \pi} \cdot f_y$. Accordingly, we abbreviate $\wh{\pi}_i = \pi_i(\wh{\theta})$, $i=1,\ldots,n$, 
and $\wh{\M{W}} = \text{diag}(1- \wh{\pi}_1, \ldots, 1- \wh{\pi}_n)$. 

Regarding $\wh{G}$, we have 
\begin{alignat*}{2}
&\frac{\partial \ell_{i,\text{p}}(\theta)}{\partial \beta} \dev{\theta}{\wh{\theta}} &&= \frac{1}{\wh{\sigma}^2} (1 - \wh{\pi}_i) \,  \M{x}_i \, (\M{x}_i^{\T} \wh{\beta} - y_i),\\[1ex]
&\frac{\partial \ell_{i,\text{p}}(\theta)}{\partial \sigma^2} \dev{\theta}{\wh{\theta}} &&= -\frac{(1 - \wh{\pi}_i) r_i^2(\wh{\beta})}{2 \wh{\sigma}^4} + \frac{(1 - \wh{\pi}_i)}{2 \wh{\sigma}^2}, \\[1ex]
&\frac{\partial \ell_{i,\text{p}}(\theta)}{\partial \alpha} \dev{\theta}{\wh{\theta}} &&= \frac{1 - \wh{\pi}_i}{1 - \wh{\alpha}} - \frac{\wh{\pi}_i}{\wh{\alpha}}, \qquad i=1,\ldots,n, 
\end{alignat*}
where for $\beta \in \R^d$, we use $r_i(\beta) =\nnorm{y_i - \M{x}_i^{\T} \beta}_2$. Moreover, regarding $\wh{H}$, one computes
\begin{align}
&\frac{\partial^2 \ell_{\text{p}}(\theta)}{\partial \beta \partial \beta^{\T}} \dev{\theta}{\wh{\theta}} = \frac{1}{\wh{\sigma}^2} \M{X}^{\T} (\wh{\M{W}} -  \wh{\bm{\Upsilon}})\M{X}, \quad 
\wh{\bm{\Upsilon}}= \text{diag}\left(\wh{\pi}_1 (1 - \wh{\pi}_1) \frac{|y_1 - \M{x}_1^{\T} \wh{\beta}|}{\wh{\sigma}}, \ldots, 
\wh{\pi}_n (1 - \wh{\pi}_n) \frac{|y_n - \M{x}_n^{\T} \wh{\beta}|}{\wh{\sigma}}\right), \notag
\\[2ex] &\frac{\partial^2 \ell_{\text{p}}(\theta)}{\partial \beta \partial \sigma^2} \dev{\theta}{\wh{\theta}} = -\frac{1}{\wh{\sigma}^4} \M{X}^{\T} \wh{\M{W}} (\M{X} \wh{\beta} - \M{y}) + \frac{1}{\wh{\sigma}^2} \M{X}^{\T} \wh{\bm{\Psi}} (\M{X} \wh{\beta} - \M{y}), \notag \\ 
&\qquad \qquad \quad \; \; \wh{\bm{\Psi}} = \text{diag}\left(\wh{\pi}_1 (1 - \wh{\pi}_1) \left( \frac{r_1^2(\wh{\beta})}{2 \wh{\sigma}^4} - \frac{1}{2 \wh{\sigma}^2} \right), \ldots, \wh{\pi}_n (1 - \wh{\pi}_n) \left( \frac{r_n^2(\wh{\beta})}{2 \wh{\sigma}^4} - \frac{1}{2 \wh{\sigma}^2} \right) \right), \notag \\[2ex] &\frac{\partial^2 \ell_{\text{p}}(\theta)}{\partial \beta \partial \alpha} \dev{\theta}{\wh{\theta}} = \frac{1}{\wh{\sigma}^2} \M{X}^{\T} \wh{\M{W}}' (\M{X} \wh{\beta} - \M{y}), \quad \wh{\M{W}}' = \text{diag}(-\wh{\pi}_1', \ldots, -\wh{\pi}_n') \label{eq:Wprime} \\[2ex] &\frac{\partial^2 \ell_{\text{p}}(\theta)}{\partial^2 \sigma^2} \dev{\theta}{\wh{\theta}} = \su \frac{(1 - \wh{\pi}_i) r_i^2(\wh{\beta})}{(\wh{\sigma}^2)^3} - \su \frac{1}{2 \wh{\sigma}^4} (1 - \wh{\pi}_i) + \su \left( \frac{r_i^2(\wh{\beta})}{2 \wh{\sigma}^4} - \frac{1}{2 \wh{\sigma}^2} \right)^2 \wh{\pi}_i (1 - \wh{\pi}_i),
\notag \\[2ex]
&\frac{\partial^2 \ell_{\text{p}}(\theta)}{\partial \sigma^2 \partial \alpha} \dev{\theta}{\wh{\theta}} = -\su \frac{\wh{\pi}_i' r_i^2(\wh{\beta})}{2 \wh{\sigma}^4} + \frac{1}{2 \wh{\sigma}^2} \su \wh{\pi}_i', \quad \frac{\partial^2 \ell_{\text{p}}(\theta)}{\partial^2 \alpha^2} \dev{\theta}{\wh{\theta}} = \su \left( \frac{\wh{\pi}_i}{\wh{\alpha}} -  \frac{1 - \wh{\pi}_i}{1- \wh{\alpha}} \right)^2 \notag,
\end{align}
where in \eqref{eq:Wprime}, $\wh{\pi}_i' = \frac{\partial \pi_i(\theta)}{\partial \alpha} \dev{\theta}{\wh{\theta}} = -\frac{1 - \wh{\pi}_i}{1 - \wh{\alpha}} + (1 - \wh{\pi}_i)(\frac{1 - \wh{\pi}_i}{1 - \wh{\alpha}} - \frac{\wh{\pi}_i}{\wh{\alpha}}), \; i=1,\ldots,n$. 
\section{Expressions for the gradient and Hessian in $\S$\ref{subsec:simultaneous}}\label{app:gradhess}
We here provide expressions for $g^{(k)}$ and $F^{(k)}$ in the 
Fisher scoring step \eqref{eq:fisher_scoring_step}. The gradient
of the expected complete data likelihood \eqref{eq:completedata_nodecoupling} is given by
{\small \begin{equation*}
g^{(k)} =  \left[ \begin{array}{l} \frac{1}{\sigma^{2(k)}} \left( \M{X}^{\T} \M{W}^{(k)} \M{X} \beta^{(k)} - \M{X}^{\T} \M{W}^{(k)} \M{y} \right) - \\[1ex]
\qquad \qquad \qquad\qquad  \qquad\qquad \qquad -\left( \su \frac{\pi_i^{(k)} y_i^2}{(\sigma^{2(k)} + \nnorm{\beta^{(k)}}_2^2)^2} - \su  \pi_i^{(k)} \frac{1}{(\sigma^{2(k)} + \nnorm{\beta^{(k)}}_2^2)} \right) \beta^{(k)}  \\[7.5ex]
 -\su (1 - \pi_i^{(k)}) \frac{r_i^2(\beta^{(k)})}{2(\sigma^{2(k)})^2} + \su (1- \pi_i^{(k)}) \frac{1}{2} \frac{1}{\sigma^{2(k)}} - \\[2.5ex]
 \qquad \qquad \qquad\qquad \qquad \qquad \qquad - \su \pi_i^{(k)} \frac{y_i^2}{2 (\sigma^{2(k)} + \nnorm{\beta^{(k)}}_2^2)^2} + \su \frac{1}{2} \pi_i^{(k)} \frac{1}{(\sigma^{2(k)} + \nnorm{\beta^{(k)}}_2^2)} \end{array} \right],
\end{equation*}}
with $r_i(\beta)$ as defined in Appendix \ref{app:sandwich}. The matrix $F^{(k)}$ can be calculated as 
\begin{equation*}
\left[ \begin{array}{cc} \frac{1}{\sigma^{2(k)}} \M{X}^{\T} \M{W}^{(k)}    \M{X} + 2 \su \frac{\pi_i^{(k)}}{\sigma^{2(k)} + \nnorm{\beta^{(k)}}_2^2} \beta^{(k)} \beta^{(k)\T} & \qquad \beta^{(k)} \su \frac{\pi_i^{(k)}}{\sigma^{2(k)} + \nnorm{\beta^{(k)}}_2^2} \\[2.5ex]
\beta^{(k)\T} \su \frac{\pi_i^{(k)}}{\sigma^{2(k)} + \nnorm{\beta^{(k)}}_2^2} & \su (1 - \pi_i^{(k)}) \frac{1}{2 (\sigma^{2(k)})^2} + \su \pi_i^{(k)} \frac{1}{2 (\sigma^{2(k)} + \nnorm{\beta^{(k)}}_2^2)}
\end{array} \right].
\end{equation*}
\end{document}